\documentclass[notitlepage,nofootinbib,preprintnumbers,amssymb,superscriptaddress,eqsecnum]{revtex4-1}
\usepackage{amsfonts,amssymb,mathtools,graphicx,color,bm}
\definecolor{ultramarine}{rgb}{0.07, 0.04, 0.56}
\definecolor{cadmiumgreen}{rgb}{0.0, 0.42, 0.24}
\definecolor{indigo(dye)}{rgb}{0.0, 0.25, 0.42}
\usepackage[linktocpage=true,breaklinks]{hyperref}
\hypersetup{
colorlinks=true,
citecolor=ultramarine,
linkcolor=cadmiumgreen,
urlcolor=indigo(dye),
}

\usepackage{autobreak}
\usepackage{cancel}
\newcommand{\fr}[2]{\frac{#1}{#2}}
\newcommand{\pa}{\partial}

\newcommand{\na}{\nabla}
\newcommand{\bra}[1]{\left( #1 \right)}  
\newcommand{\brb}[1]{\left[ #1 \right]}  
\newcommand{\brc}[1]{\left\{ #1 \right\}}  
\newcommand{\be}{\begin{equation}}  
\newcommand{\ee}{\end{equation}}
\newcommand{\bem}{\begin{bmatrix}}
\newcommand{\eem}{\end{bmatrix}}
\newcommand{\Mpl}{M_{\rm Pl}}

\newcommand{\la}{\lambda}
\newcommand{\si}{\sigma}

\newcommand{\mn}{{\mu \nu}}
\newcommand{\mC}{\mathcal{C}}
\newcommand{\mE}{\mathcal{E}}
\newcommand{\mK}{\mathcal{K}}

\newcommand{\mO}{\mathcal{O}}

\begin{document}

\preprint{KOBE-COSMO-20-05, YITP-20-35}

\title{General Relativity solutions with stealth scalar hair in quadratic higher-order scalar-tensor theories}

\author{Kazufumi Takahashi}
\affiliation{Department of Physics, Kobe University, Kobe 657-8501, Japan}

\author{Hayato Motohashi}
\affiliation{Division of Liberal Arts, Kogakuin University, 2665-1 Nakano-machi, Hachioji, Tokyo, 192-0015, Japan}
\affiliation{Center for Gravitational Physics, Yukawa Institute for Theoretical Physics, Kyoto University, Kyoto 606-8502, Japan}

\begin{abstract}
We explore General Relativity solutions with stealth scalar hair in general quadratic higher-order scalar-tensor theories.
Adopting the assumption that the scalar field has a constant kinetic term, we derive in a fully covariant manner a set of conditions under which the Euler-Lagrange equations allow General Relativity solutions as exact solutions in the presence of a general matter component minimally coupled to gravity.
The scalar field possesses a nontrivial profile, which can be obtained by integrating the condition of constant kinetic term for each metric solution.
We demonstrate the construction of the scalar field profile for several cases including the Kerr-Newman-de Sitter spacetime as a general black hole solution characterized by mass, charge, and angular momentum in the presence of a cosmological constant.
We also show that asymptotically anti-de Sitter spacetimes cannot support nontrivial scalar hair.
\end{abstract}

\maketitle

\section{Introduction}\label{sec:intro}

The direct detection of gravitational waves from a binary black hole merger opened the era of testing gravity at strong-field and dynamical regimes~\cite{Abbott:2016blz}.  
So far, observations are consistent with the Kerr spacetime in General Relativity (GR), but it does not necessarily mean that only GR is the allowed theory, since theories of modified gravity can in general allow spacetime solutions of the same (or a similar) form as those in GR (see e.g.\ \cite{Stephani:2003tm,Griffiths:2009dfa}).
Hence, it is important to clarify what kind of modified gravity theories allow such solutions and how to distinguish them from GR with precise observational data available now or promised in the near future.

As a simple framework of modified gravity, scalar-tensor theories possess an additional scalar degree of freedom, characterizing the deviation from GR.  
In general, a metric solution of a scalar-tensor theory is accompanied by a nontrivial scalar field profile.
The simplest or trivial one is the constant profile, for which a set of conditions guaranteeing the existence of 
GR solutions was derived in a fully covariant manner for general higher-derivative scalar-tensor theories~\cite{Motohashi:2018wdq}.  In this case, since terms with derivatives of the scalar field in the Euler-Lagrange (EL) equations vanish after plugging 
the constant scalar field profile, the existence condition was derived for a general theory with multiple scalar fields with arbitrarily higher-order derivatives in a $D$-dimensional spacetime in the presence of a general matter component.

To deal with a more general ansatz for the scalar field profile, one may restrict the action of higher-derivative theories.  For the construction of sensible higher-derivative theories, an important guiding principle is the absence of Ostrogradsky ghost~\cite{Woodard:2015zca}, namely, extra degree(s) of freedom associated with higher-order derivatives, whose quantum nature has also been studied~\cite{Raidal:2016wop,Smilga:2017arl,Motohashi:2020psc}.
The Ostrogradsky ghost can be evaded by imposing a certain set of degeneracy conditions~\cite{Motohashi:2014opa,Langlois:2015cwa,Motohashi:2016ftl,Klein:2016aiq,Motohashi:2017eya,Motohashi:2018pxg}, which can be systematically obtained by Hamiltonian analysis.
Upon the degeneracy condition, degenerate higher-order scalar-tensor (DHOST) theories have been constructed for those with quadratic/cubic order in the second derivative of the scalar field~\cite{Langlois:2015cwa,Crisostomi:2016czh,BenAchour:2016fzp} as well as further generalizations~\cite{Takahashi:2017pje,Langlois:2018jdg}, which serve as a general framework of healthy scalar-tensor theories with higher derivatives.\footnote{By requiring the absence of Ostrogradsky ghost only in the unitary gauge, one can obtain even broader frameworks~\cite{Gao:2014soa,DeFelice:2018mkq,Gao:2018znj,Motohashi:2020wxj}.}
Indeed, it includes Horndeski~\cite{Horndeski:1974wa,Nicolis:2008in,Deffayet:2009wt,Deffayet:2009mn,Deffayet:2011gz,Kobayashi:2011nu} and Gleyzes-Langlois-Piazza-Vernizzi (GLPV) theories~\cite{Gleyzes:2014dya,Gleyzes:2014qga} as a special subclass (see \cite{Langlois:2018dxi,Kobayashi:2019hrl} for recent reviews).

As a more general scalar field configuration, the ansatz of constant kinetic term would be the next simplest one.  
In this case, while the metric is the same as in GR, the scalar field is allowed to have a nontrivial profile, meaning that it is hidden or stealth.
Stealth solutions with a constant kinetic term in DHOST theories have been investigated for the Schwarzschild--(anti-)de Sitter spacetime~\cite{BenAchour:2018dap,Motohashi:2019sen,Minamitsuji:2019shy} as well as the Kerr-de Sitter spacetime~\cite{Charmousis:2019vnf}, generalizing the known stealth solutions of the same type in the Horndeski~\cite{Babichev:2013cya,Kobayashi:2014eva} and GLPV~\cite{Babichev:2016kdt,Babichev:2017guv,Babichev:2017lmw} theories.
Perturbations around the stealth solutions have also been extensively investigated~\cite{Ogawa:2015pea,Takahashi:2015pad,Takahashi:2016dnv,Tretyakova:2017lyg,Babichev:2017lmw,Babichev:2018uiw,Takahashi:2019oxz,deRham:2019slh,Charmousis:2019fre,Motohashi:2019ymr}.
In particular, the stability condition for the odd-parity perturbations around a general static spherically symmetric vacuum spacetime was clarified in~\cite{Takahashi:2019oxz}.
Also, perturbation analyses clarified that some class of stealth solutions 
exhibits strong coupling~\cite{Babichev:2018uiw,Minamitsuji:2018vuw,deRham:2019slh}, among which the one with a timelike scalar field derivative can be cured by a weak and controlled violation of the degeneracy condition dubbed ``scordatura''~\cite{Motohashi:2019ymr}, named after an Italian word meaning ``detuning.'' 
It is natural to expect such a detuning from the effective-field-theory (EFT) point of view. 
It renders the strong coupling scale sufficiently high, whereas the Ostrogradsky ghost associated with the violation of degeneracy conditions is adjusted to show up only above the EFT cutoff scale.
Since the scordatura mechanism allows a wider class of weakly-coupled stealth solutions, it would be intriguing to investigate stealth solutions in higher-derivative theories from a more general perspective,  
i.e., broadening the target theories and metric solutions, and also taking into account the presence of a matter component.

So far, most stealth solutions with the ansatz of constant kinetic term in DHOST theories have been considered in vacuum (however, see \cite{Babichev:2015rva}, where a charged black hole solution with stealth scalar hair was obtained in a subclass of Horndeski theory with a nonminimally coupled gauge field). 
In principle, by using the conformal/disformal transformation, an exact solution in DHOST theories can be recast into another exact solution in a simpler framework, e.g., the Horndeski theory~\cite{Achour:2016rkg}.
Thus, for vacuum solutions, it might be more reasonable to consider exact solutions in simpler theories.
However, if one takes into account the coupling to a matter component, performing a metric redefinition implies a change of frames, and hence an exact solution in a DHOST theory minimally coupled to the matter field is in general transformed into an exact solution in another DHOST theory with a nontrivial matter coupling.
In other words, the presence of a matter component prescribes the notion of frames and then the higher-derivative interactions in the action and the exact solutions equip more solid meanings.
Hence, it is interesting to look for stealth solutions in the presence of a matter component.

In this paper, we generalize the analysis of \cite{Motohashi:2018wdq} to the case with the ansatz of constant kinetic term and explore exact solutions in the presence of a general matter component in higher-order scalar-tensor theories in a covariant manner.  In \S\ref{sec:cov}, we derive covariant EL equations for general higher-derivative theories and clarify a set of conditions under which metrics satisfying the Einstein equation in GR with a matter component are allowed as exact solutions.  Our condition thus applies to any GR solution, for each of which the scalar field profile can be derived by integrating the condition of constant kinetic term.  In \S\ref{sec:sca}, we demonstrate the construction of the scalar field profile for several specific cases.  In particular, we investigate the Kerr--Newman--(anti-)de Sitter 
spacetime as a general black hole solution characterized by mass, charge, and angular momentum in the presence of a cosmological constant.  In \S\ref{sec:con}, we draw our conclusions.

\section{Existence conditions for General Relativity solutions}\label{sec:cov}

In this section, we derive covariant EL equations and clarify a set of conditions for them to allow GR metric solutions as exact solutions.
We consider a class of theories described by the following action with at most quadratic order of second derivatives of the scalar field: 
	\be
	S=\int d^4x\sqrt{-g}\brb{F_0+F_1\Box\phi+F_2R+\sum_{I=1}^{5}A_IL_I^{(2)}} + \int d^4x\sqrt{-g}L_m, \label{qHOST}
	\ee
where the coupling functions $F_0$, $F_1$, $F_2$, and $A_I$ ($I=1,\cdots,5$) are functions of $\phi$ and $X\coloneqq \phi_\mu\phi^\mu$, and 
	\be
	L_1^{(2)}\coloneqq \phi^{\mn}\phi_{\mn},~~~L_2^{(2)}\coloneqq (\Box\phi)^2,~~~L_3^{(2)}\coloneqq \phi^\mu\phi_{\mn}\phi^\nu\Box\phi,~~~L_4^{(2)}\coloneqq \phi^\mu\phi_{\mn}\phi^{\nu\la}\phi_\la,~~~L_5^{(2)}\coloneqq (\phi^\mu\phi_{\mn}\phi^\nu)^2,
	\ee 
with $\phi_\mu\coloneqq \na_\mu\phi$ and $\phi_{\mn}\coloneqq \na_\mu\na_\nu\phi$.
If the coupling functions depend on $X$ only, the action respects the shift symmetry~$\phi\to \phi+{\rm const}$.
If we choose the functions~$A_2$, $A_4$, and $A_5$ as
	\be
	\begin{split}
	A_2&=-A_1\ne -\fr{F_2}{X}, \\
	A_4&=\fr{1}{8(F_2-XA_1)^2}\bigl\{4F_2\brb{3(A_1-2F_{2X})^2-2A_3F_2}-A_3X^2(16A_1F_{2X}+A_3F_2) \\
	&~~~~~~~~~~~~~~~~~~~~~~~~+4X\bra{3A_1A_3F_2+16A_1^2F_{2X}-16A_1F_{2X}^2-4A_1^3+2A_3F_2F_{2X}}\bigr\}, \\
	A_5&=\fr{1}{8(F_2-XA_1)^2}(2A_1-XA_3-4F_{2X})\brb{A_1(2A_1+3XA_3-4F_{2X})-4A_3F_2},
	\end{split} \label{DC}
	\ee
the action~\eqref{qHOST} reduces to the one for the class Ia of quadratic DHOST theories~\cite{Langlois:2015cwa}.
Here, a subscript~$X$ denotes a derivative with respect to $X$.
For generality, we do not impose these degeneracy conditions unless otherwise stated.

For the matter Lagrangian $L_m$, we assume that the matter component minimally couples to gravity. 
Then, the equations of motion for the matter sector remain the same as those in GR regardless of the scalar field interactions.
Therefore, in the following we shall focus on the EL equations for the metric and the scalar field.

Now we assume a solution with $X=X_0={\rm const}$, 
which allows us to express higher-derivative terms of $\phi$ as 
    \begin{align}
    \begin{split}
    \phi^\mu\phi_\mn&=0, \\
    \phi^\la\na_\la\phi_\mn&=-R_{\mu\la\nu\si}\phi^\la\phi^\si-\phi_\mu^\la\phi_{\la\nu}, \\
    \phi^\la\na_\la\Box\phi&=-R_{\la\si}\phi^\la\phi^\si-\phi_{\alpha\beta}^2,
    \end{split}
    \end{align}
where $\phi_{\alpha\beta}^n$ denotes a scalar quantity defined by $\phi_{\alpha\beta}^n\coloneqq \phi_{\alpha_1}^{\alpha_2}\phi_{\alpha_2}^{\alpha_3}\cdots\phi_{\alpha_{n-1}}^{\alpha_n}\phi_{\alpha_n}^{\alpha_1}$.
Then, the EL equation for the metric is given by $\mE_\mn=0$, with 
\begin{align}
\mE_\mn=& 2F_2 R_{\mu\nu} \nonumber \\
    &-\brc{F_0+R F_2 - X_0 (F_{1\phi} + 2 F_{2\phi\phi})+A_1\phi_{\alpha\beta}^2 -A_2\brb{(\Box\phi)^2-2\phi_{\alpha\beta}^2- 2\phi^\alpha \phi^\beta R_{\alpha\beta}  } 
    -2(F_{2\phi}+X_0A_{2\phi})\Box\phi  }g_\mn \nonumber \\
    &+\bigl\{2(F_{0X}+R F_{2X}-F_{1\phi} - F_{2\phi\phi} ) + (2F_{1X} - 4A_{2\phi} - X_0A_{3\phi} )\Box\phi \nonumber \\
    &\quad +A_3 \phi^\alpha\phi^\beta R_{\alpha\beta}  -(A_3+2A_{1X})\brb{(\Box\phi)^2-\phi_{\alpha\beta}^2}
    +2(A_{1X}+A_{2X})(\Box\phi)^2\bigr\}\phi_\mu\phi_\nu \nonumber \\
    &+2A_1\brb{ (\Box\phi)\phi_\mn - \phi_\mu^\la\phi_{\la\nu} - R_{\mu\la\nu\si}\phi^\la\phi^\si} 
    +4A_2R_{\la(\mu}\phi_{\nu)}\phi^\la- 2\phi_{\mu\nu} ( F_{2\phi} - X_0 A_{1\phi} ) \nonumber \\
    &-4(A_1+A_2)\phi_{(\mu}\Box \phi_{\nu)}- T_{\mu\nu}, \label{EmnHOST}
\end{align}
where the coupling functions and their derivatives are evaluated at $X=X_0$.
Here, a subscript~$\phi$ denotes a derivative with respect to $\phi$ and $T_{\mu\nu} \coloneqq -\fr{2}{\sqrt{-g}} \fr{\delta(\sqrt{-g}L_m)}{\delta g^{\mu\nu}}$ is the stress-energy tensor for the matter sector.
It should be noted that, unless $\phi={\rm const}$, the EL equation for the scalar field, which we denote by $\mE_\phi=0$, is automatically satisfied for any configuration~$(g_\mn,\phi)$ that satisfies $\mE_\mn=0$ and the equations of motion for the matter field thanks to the Noether identity
associated with general covariance (see \cite{Motohashi:2016prk} for related discussions).
In other words, $\mE_\phi=0$ can be reproduced from other EL equations and hence is a redundant equation.
Note also that the terms with $A_4$ and $A_5$ do not contribute to the EL equations under the condition $X={\rm const}$.
Equation~\eqref{EmnHOST} allows GR solutions satisfying $G_{\mu\nu} = \Mpl^{-2} T_{\mu\nu} - \Lambda g_{\mu\nu}$ with the reduced Planck mass~$\Mpl=(8\pi G)^{-1/2}$ if the following conditions are satisfied at $X=X_0$: 
\begin{align} 
\begin{split}
&F_0+2\Lambda F_2 - X_0 (F_{1\phi} + 2 F_{2\phi\phi}) = 0,\quad
2F_{1X} + 4A_{1\phi} - X_0A_{3\phi}=0, \quad
F_{2\phi} - X_0 A_{1\phi} =0 , \\
&A_1 = 0, \quad
A_2 = 0, \quad
A_{1\phi}+A_{2\phi} = 0, \quad
A_{1X}+A_{2X} = 0, \quad
A_3+2A_{1X} = 0 , \quad
(2\Mpl^{-2}F_2 - 1) T_{\mu\nu} = 0, \\
&F_{0X}-F_{1\phi} - F_{2\phi\phi} + \Lambda (4 F_{2X} - X_0 A_{1X}) = \Mpl^{-2}[ F_{2X}T + A_{1X} (\phi^\alpha\phi^\beta T_{\alpha\beta}-TX_0/2) ],
\end{split}
    \label{exist_cond_HOST}
\end{align}
with $T\coloneqq T^\mu_\mu$.
If we restrict ourselves to DHOST theories satisfying $A_2(\phi,X)=-A_1(\phi,X)$, the terms proportional to the sum~$A_1+A_2$ and its derivatives in the EL equation~\eqref{EmnHOST} vanish.
As a result, we obtain a simpler set of conditions for the existence of GR solutions, which amounts to \eqref{exist_cond_HOST} with the replacement~$A_2\to -A_1$, $A_{2\phi}\to -A_{1\phi}$, etc.

An example of theories that can trivially satisfy the conditions~\eqref{exist_cond_HOST} for an arbitrary matter component is shift-symmetric DHOST theories with the following coupling functions:
    \be
    F_1=A_1=A_2=0,\quad
    F_2=\fr{\Mpl^2}{2},\quad
    A_3=\mu F_{0X},\quad
    A_4=-\mu F_{0X}\bra{1+\fr{\mu}{4\Mpl^2}X^2F_{0X}},\quad
    A_5=\bra{\fr{\mu}{\Mpl}}^2XF_{0X}^2,
    \ee
where $\mu$ is constant.
In this case, the conditions~\eqref{exist_cond_HOST} read
    \be
    F_0(X_0)+\Mpl^2\Lambda=0,\quad
    F_{0X}(X_0)=0,
    \ee
where the second equation fixes the value of $X_0$ algebraically, and then the first equation determines the value of $\Lambda$.

There are several remarks along the same lines as the sufficient conditions for GR solutions with $\phi={\rm const}$ for arbitrarily higher-order scalar-tensor theories obtained in \cite{Motohashi:2018wdq}.
As mentioned above, $A_4$ and $A_5$ do not contribute to the EL equations for 
solutions with $X={\rm const}$, and hence they are not constrained by the above conditions.
Therefore, these conditions are independent of or consistent with the degeneracy conditions \eqref{DC} on $A_4$ and $A_5$.
This is similar to the situation for the conditions for GR solutions with a constant scalar field profile derived in \cite{Motohashi:2018wdq} for theories with arbitrarily higher-order derivatives, which are independent of the degeneracy conditions with arbitrarily higher-order derivatives~\cite{Motohashi:2017eya,Motohashi:2018pxg}.
Another caveat is that the above set of conditions is a sufficient condition for the covariant EL equations to allow GR solutions satisfying $G_{\mu\nu} = \Mpl^{-2} T_{\mu\nu} - \Lambda g_{\mu\nu}$ with $X={\rm const}$, and hence there would be a weaker condition for a particular metric ansatz and/or a matter profile.

As a specific example, let us consider a more restricted case with the vacuum Einstein manifold~$G_\mn=-\Lambda g_\mn$ in shift-symmetric DHOST theories, i.e., those with the coupling functions depending only on $X$.
In this case, the left-hand side of the EL equation for the metric is given by
    \begin{align}
    \mE_\mn=&
    -\brc{F_0+2\Lambda F_2+A_1\brb{(\Box\phi)^2-\phi_{\alpha\beta}^2-2\Lambda X_0 } }g_\mn \nonumber \\
    &+\brc{2F_{0X}+\Lambda (8 F_{2X}-4 A_1+ X_0 A_3 ) + 2F_{1X}\Box\phi -(A_3+2A_{1X})\brb{(\Box\phi)^2-\phi_{\alpha\beta}^2} }\phi_\mu\phi_\nu \nonumber \\
    &+2A_1[ (\Box\phi)\phi_\mn- \phi_\mu^\la\phi_{\la\nu}- R_{\mu\la\nu\si}\phi^\la\phi^\si ]. \label{EmnDHOST}
    \end{align}
This equation can be satisfied if the following conditions hold at $X=X_0$:\footnote{Focusing on static spherically symmetric vacuum solutions in DHOST theories, the condition~\eqref{solcon} is weaker than the sufficient condition for strong coupling obtained in \cite{deRham:2019slh}. 
Thus, perturbations around a GR solution may or may not be strongly coupled, for the former of which one can employ the scordatura technique~\cite{Motohashi:2019ymr}.}
\begin{align} \label{solcon}
F_0+2\Lambda F_2 = 0 ,\quad
F_{0X}+\Lambda (4 F_{2X}- X_0 A_{1X})= 0,\quad
F_{1X}=0, \quad
A_1 = 0, \quad
A_3+2A_{1X} = 0.
\end{align}
These conditions actually coincide with those of the Case~2-$\Lambda$ given by Eq.~(43) in \cite{Motohashi:2019sen} for the Schwarzschild--(anti-)de Sitter spacetime.
The conditions~\eqref{solcon} also recover those given in \cite{Charmousis:2019vnf} for the subclass with $A_1(X)=A_2(X)=F_1(X)=0$.
However, under a more restricted setup, one would obtain a weaker set of conditions.
Indeed, for the case of the Schwarzschild--(anti-)de Sitter spacetime with a linearly time-dependent scalar field~$\phi=qt+\psi(r)$ having $X=-q^2$, one can verify the following identity:
    \be
    (\Box\phi)\phi_\mn- \phi_\mu^\la\phi_{\la\nu}- R_{\mu\la\nu\si}\phi^\la\phi^\si
    =\Lambda(q^2 g_\mn +\phi_\mu\phi_\nu). \label{id}
    \ee
This identity can be used to reduce \eqref{EmnDHOST} as
    \be
    \mE_\mn=
    -\brb{F_0+2\Lambda (F_2-X_0 A_1)}g_\mn
    +\brb{2F_{0X}+\Lambda (8 F_{2X}-2 A_1+4X_0 A_{1X}+ 3X_0 A_3)+2F_{1X}\Box\phi}\phi_\mu\phi_\nu.
    \ee
Thus, $\mE_\mn$ vanishes if
    \be
    F_0+2\Lambda (F_2-X_0 A_1)=0,\quad
    2F_{0X}+\Lambda (8 F_{2X}-2 A_1+4X_0 A_{1X}+ 3X_0 A_3)=0,\quad
    F_{1X}=0, \label{reduced_cond}
    \ee
which is weaker than \eqref{solcon}.
The first two conditions in \eqref{reduced_cond} reproduce Eq.~(25) in \cite{Takahashi:2019oxz}, which was derived by reducing the higher-order EL equations to second-order differential equations for a general static spherically symmetric metric.
Moreover, it is consistent with the Case~1-$\Lambda$ given by Eq.~(42) in \cite{Motohashi:2019sen}.
As a specific example, one can check that the self-tuned Schwarzschild-de Sitter solution found in \cite{Babichev:2013cya} satisfies the condition~\eqref{reduced_cond}.

Likewise, for the case of trivial or constant scalar profile~$\phi=\phi_0={\rm const}$, the EL equations for the metric and the scalar field are simplified as
\begin{align}
    \mE_\mn=& 2F_2 R_{\mu\nu}-(F_0+R F_2)g_\mn -  T_{\mu\nu}=0, \\
    \mE_\phi=&F_{0\phi} + F_{2\phi}R=0.
\end{align}
Note that here we need the EL equation for $\phi$ as it cannot be reproduced from other EL equations when $\phi={\rm const}$.
Consequently, the sufficient condition for the existence of GR solutions is much relaxed:
\begin{align}
    F_0+2\Lambda F_2 = 0 ,\quad
    F_{0\phi}+4\Lambda F_{2\phi}= 0,\quad
    (2\Mpl^{-2}F_2-1)T_\mn=0,\quad
    F_{2\phi} T =0, \label{exist_cond_phi_const}
\end{align}
where the coupling functions are evaluated at $(\phi,X)=(\phi_0,0)$.
This is consistent with the condition derived in \cite{Motohashi:2018wdq} for general multi-scalar higher-derivative theories in $D$-dimensional spacetime.
The conditions~\eqref{exist_cond_phi_const} are trivially satisfied for theories with $F_2=\Mpl^2/2$.
In this case, the conditions read
    \be
    F_0(\phi_0,0)+\Mpl^2\Lambda=0,\quad
    F_{0\phi}(\phi_0,0)=0,
    \ee
where the second equation fixes the value of $\phi_0$, and then the first equation determines the value of $\Lambda$.

While we need to keep in mind the above caveats, an advantage of the above sufficient conditions also originates from their fully covariant derivation. 
If the coupling functions satisfy the condition at $X=X_0$, any 
GR solution in the presence of a matter component is allowed as an exact background solution.
As shown in the \hyperref[App]{Appendix}, it is straightforward to obtain a similar set of conditions for an arbitrary spacetime dimension~$D$.
Also, there should exist a similar identity as \eqref{id} in $D$ dimensions under some restricted setup, which applies to known solutions, e.g., in~\cite{Bravo-Gaete:2014haa} obtained in three dimensions.

\section{Scalar field profile}\label{sec:sca}

In \S\ref{sec:cov}, we derived a set of sufficient conditions for the existence of GR solutions in the presence of a matter field.
Since the derivation was performed in a covariant manner, any GR metric is allowed as an exact solution.
For each metric, we choose an appropriate ansatz for the scalar field and derive its profile by integrating the condition~$X=X_0$. 
In this section, we consider several specific spacetimes and demonstrate the construction of the scalar field profile.
Note that the construction solely relies on the condition~$X=X_0$ and the form of the metric.
As we discussed above, since the set of conditions derived in \S\ref{sec:cov} is a sufficient condition, the metrics discussed below could also be a solution of theories that are not covered by those obtained in \S\ref{sec:cov}.
Indeed, it would be possible to consider GR solutions in theories beyond our action~\eqref{qHOST} containing at most quadratic interactions of second derivatives of $\phi$.
Nevertheless, the scalar field profiles obtained in this section apply to even such cases so long as the ansatz of constant kinetic term is satisfied.

\subsection{Static spherically symmetric spacetime}\label{ssec:sss}

Let us consider a static spherically symmetric spacetime described by the metric, 
\be \label{sss-metric-ansatz} g_{\mu\nu}dx^\mu dx^\nu = - A(r)dt^2 + \fr{dr^2}{B(r)} + r^2(d\theta^2+\sin^2\theta d\varphi^2) . \ee
For the scalar field profile, we take a linearly time-dependent profile, 
\be \label{sss-scalar-ansatz} \phi(t,r) = q t + \psi(r) , \ee
with $q$ being constant.
With these ansatzes, the kinetic term of the scalar field is given by
\be X(r) = -\fr{q^2}{A} + B \psi'^2, \ee
where a prime denotes a derivative with respect to $r$. 
From the condition~$X=X_0$, we obtain 
\be \label{diff-psi} \psi'=\pm \sqrt{\fr{q^2+X_0A}{AB}}, \ee
where we assume $\fr{q^2+X_0A}{AB} \geq 0$.
For a given set~$(A(r), B(r))$, one can obtain $\psi(r)$ by integrating \eqref{diff-psi}.

In general, one can make $\psi(r)$ regular at event horizon(s) by choosing the appropriate branch.
Let us suppose that the spacetime has a black hole horizon at $r=r_b$ and/or a cosmological horizon at $r=r_c$.  
Around the black hole or cosmological horizon, which we simply denote by $r=r_h$,
we assume $A(r)=A_h (r-r_h) + \mO(|r-r_h|^2)$ and $B(r)=B_h (r-r_h) + \mO(|r-r_h|^2)$ with $A_h$ and $B_h$ being nonvanishing constants.
Then, one can integrate \eqref{diff-psi} to obtain
\be \label{prs} \psi = \pm q r_* +\mO(|r-r_h|), \ee
where $r_* = \int dr (AB)^{-1/2} \simeq \pm (A_hB_h)^{-1/2} \log \left| \fr{r-r_h}{r_h} \right|$ is the tortoise coordinate, with the plus/minus sign for $r\gtrsim r_b$ or $r\lesssim r_c$, namely, just outside the black hole horizon~$r=r_b$ or just inside the cosmological horizon~$r=r_c$. 
Thus, the scalar field seems to be divergent at the horizons, but one can make it finite by choosing an appropriate coordinate system.
Indeed, we obtain $\phi\simeq qv$ for the plus branch of \eqref{prs} and $\phi\simeq qu$ for the minus branch,
where $v\coloneqq t+r_*$ and $u\coloneqq t-r_*$ are the advanced and retarded null coordinates, respectively.
Let us first focus on a region just outside the black hole horizon.
Using the ingoing Eddington-Finkelstein coordinates~$(v, r)$, we see that the scalar field is finite at the future event horizon for the plus branch, while it diverges for the minus branch.
Likewise, using the outgoing Eddington-Finkelstein coordinates~$(u, r)$, we see that the opposite is the case at the past event horizon.
On the other hand, for a region just inside the cosmological horizon, using the ingoing (outgoing) Eddington-Finkelstein coordinates, we can see that the scalar field is finite at the past (future) event horizon if we choose the plus (minus) branch of \eqref{prs}.
This implies that the scalar field is finite at both the horizons for an observer free-falling from the cosmological horizon to the black hole horizon or 
traveling from the black hole horizon to the cosmological horizon. 
Note that the situation is completely different when $q=0$.
In this case, we have
    \be
    \psi=\pm \int dr\sqrt{\fr{X_0}{B}}
    \simeq \pm 2\sqrt{\fr{X_0(r-r_h)}{B_h}},
    \ee
near the horizons, and thus the scalar field is finite from the beginning.

While the existence of GR solution is guaranteed by the condition~\eqref{exist_cond_HOST}, the evolution of perturbations needs to be studied.
For instance, in shift- and reflection-symmetric quadratic DHOST theories, the stability condition for the odd-parity perturbations around a static spherically symmetric vacuum spacetime with the scalar field profile~\eqref{sss-scalar-ansatz} is given by~\cite{Takahashi:2019oxz}
\be F_2>0, \quad F_2-XA_1>0, \quad F_2-\left( \frac{q^2}{A} + X \right) A_1>0, \ee
which should be maintained in addition to the existence condition~\eqref{exist_cond_HOST}.
Of course, the stability of the even-parity perturbations should also be taken into account.
However, for the even-parity sector, while the stability condition for a static scalar field background was clarified for Horndeski~\cite{Kobayashi:2014wsa} and GLPV subclasses~\cite{Mironov:2018uou}, the one for $\phi=qt+\psi(r)$ in generic DHOST theories has not been clarified yet.

Let us derive scalar field profiles for specific metrics.
First, for the Schwarzschild spacetime, 
\be A(r)=B(r)=1-\fr{r_g}{r} \eqqcolon f_{\rm S}(r), \ee 
where $r_g\coloneqq 2GM$ is the Schwarzschild radius with 
mass $M$ in the natural unit $c=1$, we obtain~\cite{Motohashi:2019sen}
\begin{align} 
\psi &= \pm \int dr \fr{ \sqrt{q^2 + X_0 f_{\rm S} } }{ f_{\rm S} } \notag\\
&= \pm\left[ r\sqrt{Q_{\rm S}} - 2qr_g {\rm artanh} \frac{q}{\sqrt{Q_{\rm S}}} + \frac{(2q^2+X_0)r_g}{\sqrt{Q_0}} {\rm artanh} \frac{\sqrt{Q_{\rm S}}}{\sqrt{Q_0} } \right] + {\rm const} , \label{psi-Sch}
\end{align}
where $Q_0\coloneqq q^2+X_0$ and $Q_{\rm S}\coloneqq q^2+X_0f_{\rm S}$.
When $X_0=-q^2$, we have 
    \be \label{psi-Sch-q}
    \psi=\pm2q\bra{\sqrt{r_gr}-r_g{\rm artanh}\sqrt{\fr{r_g}{r}}} + {\rm const}.
    \ee
The apparent difference from the expression in \cite{Motohashi:2019sen} can be absorbed into a redefinition of the integration constant.
Also, note that the expression \eqref{psi-Sch} should be understood as a representative form.
For a real~$x$, ${\rm artanh}\, x$ is real if and only if $|x|<1$, and it identically holds that ${\rm artanh}\, x = {\rm artanh} (1/x)$ up to an imaginary constant which depends on the branch choice. 
Therefore, ${\rm artanh}\, x$ or ${\rm artanh} (1/x)$ can be a solution of a same differential equation and one should choose the real one.
For instance, when $\frac{q}{\sqrt{Q_{\rm S}}}>1$, one should reinterpret the second term in the brackets of \eqref{psi-Sch} as ${\rm artanh} \frac{\sqrt{Q_{\rm S}}}{q}$.

As stressed above, our analysis applies to any GR solution in the presence of a general matter component minimally coupled to gravity.
As an example of the matter sector, let us consider the source-free Maxwell action,
\be L_m = -\fr{1}{4}F^{\mu\nu}F_{\mu\nu}, \ee
where $F_{\mu\nu}\coloneqq \pa_\mu A_\nu-\pa_\nu A_\mu$.
The static spherically symmetric solution for source-free Einstein-Maxwell equations is given by the Reissner-Nordstr\"om metric, 
\be A(r)=B(r)=1-\fr{r_g}{r}+\fr{r_e^2}{r^2} \eqqcolon f_{\rm RN}(r), \ee 
where $r_e\coloneqq Ge^2$ with electric charge $e$ in the natural unit~$4\pi\epsilon_0=1$. 
There are several possible cases (though not exclusive to each other).
For the extremal case~$r_g = 2 r_e$ with $X_0=-q^2$, we obtain 
\be \psi = \pm q \brb{  \fr{ (2 r - 3 r_e) \sqrt{(2 r - r_e) r_e} }{r - r_e} - 
4 r_e  {\rm artanh} 
\sqrt{\fr{r_e}{2r-r_e}}} + {\rm const.} \ee
For the extremal case~$r_g = 2 r_e$ with $X_0\ne -q^2$, 
\begin{align} 
\psi &= \pm  \Biggl[ \fr{(r - 2 r_e) \sqrt{ Q_{\rm RN} }}{1 - r_e/r} 
- 2 q r_e {\rm artanh}\fr{q }{\sqrt{ Q_{\rm RN} }} 
+ \fr{r_e (2 q^2 + X_0) }{\sqrt{Q_0} }  {\rm artanh}\fr{ q^2 + X_0 (1 - r_e/r) }{ \sqrt{Q_0 Q_{\rm RN}} } \Biggr] + {\rm const}, \end{align}
where $Q_{\rm RN}\coloneqq q^2 + X_0 f_{\rm RN}$.
For the sub-extremal case~$r_g > 2 r_e$ with $X_0= -q^2$, 
\be \psi=  \pm 2 q \brb{  r \sqrt{1-f_{\rm RN}} 
+ \fr{r_-^2}{ r_+ - r_- } {\rm artanh}\bra{ \fr{r}{r_-} \sqrt{1-f_{\rm RN}} } 
+ (r_+ \leftrightarrow r_-)
} 
+ {\rm const}, \ee
where $r_\pm \coloneqq \bra{ r_g\pm\sqrt{r_g^2 - 4 r_e^2} }/2$, and the third term in the brackets represents a term of the same form as the second term with the permutation~$r_+ \leftrightarrow r_-$.
Finally, for the sub-extremal case~$r_g > 2 r_e$ with $X_0\ne -q^2$, we obtain
\begin{align} 
\psi &= \pm \Biggl[ r\sqrt{Q_{\rm RN}}  
+\fr{r_g (2 q^2 + X_0)}{2\sqrt{Q_0}} {\rm artanh} \fr{ q^2 + X_0 \bra{1 - \fr{r_g}{2 r}}  }{\sqrt{Q_0 Q_{\rm RN} }  }  \notag\\
&~~~~~~~~~ +\fr{ qr_-^2 }{ r_+ - r_- } {\rm artanh} \fr{ q^2 + X_0 \bra{1 - \fr{r_g}{2 r} - \fr{r^2}{2 r_-} f'_{\rm RN} }  }{ q \sqrt{ Q_{\rm RN} } }  
+ (r_+ \leftrightarrow r_-)
\Biggr] 
+ {\rm const},
\end{align}
where the last term in the brackets denotes a term of the same form as the third term with the permutation~$r_+\leftrightarrow r_-$.
These solutions, together with the one for the Schwarzschild case~\eqref{psi-Sch}, are consistent with each other up to the ambiguity explained below \eqref{psi-Sch-q} when taking the corresponding limits.

More generally, for the Schwarzschild--(anti-)de Sitter spacetime, 
\be A(r)=B(r)=1-\fr{r_g}{r}-\fr{\Lambda r^2}{3} \eqqcolon f_{\rm SdS}(r), \ee 
or the Reissner--Nordstr\"om--(anti-)de Sitter spacetime,
\be A(r)=B(r)=1-\fr{r_g}{r}+\fr{r_e^2}{r^2}-\fr{\Lambda r^2}{3} \eqqcolon f_{\rm RNdS}(r), \ee
with $\Lambda$ being a cosmological constant, $\psi(r)$ is given by the integral,
\be \label{psiSS} \psi = \pm \int dr \fr{ \sqrt{q^2 + X_0 f } }{ f } , \ee
where $f(r)=f_{\rm SdS}(r)$ or $f_{\rm RNdS}(r)$. 
In general, the value of $\Lambda$ is determined from the condition~\eqref{exist_cond_HOST} irrespective of a possibly present bare cosmological constant in $F_0$, and hence one can tune $\Lambda$ by model parameters of the theory.

In \eqref{psiSS}, we have assumed $q^2 + X_0 f\geq 0$. 
This condition is trivially satisfied when $X_0\ge 0$.
For $X_0<0$, if we define a dimensionless parameter~$\eta$ by $\eta^2 \coloneqq -q^2/X_0$, this condition reads $|\eta| \geq \sqrt{f}$.
Provided that the function~$f(r)$ takes its maximum at some finite $r=r_0$, we have $|\eta|\ge \sqrt{f(r_0)}\eqqcolon \eta_c$.
For a generic value of $\eta$ with $|\eta|>\eta_c$, one can choose either branch of $\psi$ so that the scalar field is finite at both the black hole and cosmological horizons for a specific observer as explained above.
On the other hand, for the marginal value~$|\eta|=\eta_c$, one should change the branch of $\psi$ at $r=r_0$ so that the derivative of the scalar field is smooth there~\cite{Charmousis:2019vnf}.
Specifically, for the case of the Schwarzschild-de Sitter spacetime, when $(0<)9\Lambda r_g^2/4<1$, the spacetime has the cosmological horizon~$r=r_c$ in addition to the black hole horizon~$r=r_b(<r_c)$.
In this case, we have $r_0=\left(\frac{3r_g}{2\Lambda} \right)^{1/3}$, and thus the lower bound of $|\eta|$ is given by
\be |\eta| \geq \sqrt{1-\left( \frac{9\Lambda r_g^2}{4} \right)^{1/3}} \eqqcolon \eta_c . \ee
For $|\eta|=\eta_c$, an appropriate branch change makes $\phi\simeq q(t+r_*)$ near $r=r_b$ and $\phi\simeq q(t-r_*)$ near $r=r_c$, or vice versa.
On the contrary, for the case of the Schwarzschild--anti-de Sitter spacetime, the function~$f(r)$ can take arbitrarily large values, so the condition~$q^2+X_0f\ge 0$ cannot be satisfied for $X_0<0$.
The situation remains the same even for the Reissner--Nordstr\"om--anti-de Sitter case.
Thus, the Reissner--Nordstr\"om--anti-de Sitter spacetimes with stealth scalar hair are allowed only when $X_0\ge 0$.

One can also consider a general perfect fluid with $T^\mu_\nu={\rm diag}(-\rho,P,P,P)$ as a matter component.
With the energy density $\rho(r)$ and the  
pressure $P(r)$, the time and radial components of the Einstein equation yield
\begin{align} 
\fr{(rB)'}{r^2}-\fr{1}{r^2}=-8\pi G \rho, \quad
\fr{B(rA)'}{r^2A}-\fr{1}{r^2}=8\pi G P,
\end{align}
whereas the angular components can be rewritten in the form,
\be P'=(\rho+P)\fr{B-1-8\pi G r^2P}{2rB}, \ee
which is known as the Tolman-Oppenheimer-Volkoff equation.
For a given density profile~$\rho(r)$, one can determine the metric components~$A(r),B(r)$ and the radial pressure~$P(r)$ from the above equations,  
after which the scalar field profile~$\psi(r)$ is obtained by integrating \eqref{diff-psi}.

\subsection{Stationary axisymmetric spacetime}\label{ssec:sax}

Let us consider a general stationary axisymmetric spacetime in the presence of the source-free Maxwell action minimally coupled to gravity.
In GR, the asymptotically (anti-)de Sitter solution is known as the Kerr--Newman--(anti-)de Sitter metric, which is expressed in the Boyer-Lindquist coordinates as~\cite{Carter:1970ea,Carter:1973rla}
\be
g_{\mu\nu}dx^\mu dx^\nu = 
- \frac{\Delta_r}{\Xi^2\rho^2} (dt-a\sin^2\theta d\varphi)^2 
+ \frac{\rho^2}{\Delta_r}dr^2
+ \frac{\rho^2}{\Delta_\theta}d\theta^2
+ \frac{\Delta_\theta\sin^2\theta}{\Xi^2\rho^2} \left[ adt-(r^2+a^2) d\varphi  \right]^2, \label{KNAdS}
\ee
where
\begin{align} 
\begin{split}
\rho^2 &= r^2+a^2\cos^2\theta, \quad
\Xi =1+\frac{1}{3}\Lambda a^2, \\
\Delta_r &= (r^2+a^2)\left(1-\frac{1}{3}\Lambda r^2\right) - r_g r + r_e^2, \quad
\Delta_\theta = 1+\frac{1}{3}\Lambda a^2\cos^2\theta, 
\end{split}
\end{align}
and $a$ is the angular momentum per unit mass.
This metric describes a rotating black hole with electric charge in an expanding universe.

We adopt the following ansatz for the scalar field profile: 
\be \label{KNdSansatz} \phi = -Et + L_z\varphi + f_r(r) + f_\theta(\theta),   \ee
where $E,L_z$ are constants, and $f_r(r),f_\theta(\theta)$ are independent functions.
Here, we employ the notation $q=-E$ for the reason we shall explain below.
The equation~$X=X_0$ yields a differential equation for $f_r$ and $f_\theta$, which we can separate in the form,
\be \label{HJeq}
\Delta_\theta \left( \frac{df_\theta}{d\theta}\right)^2 + \frac{\Xi^2}{\Delta_\theta} \left(\frac{L_z}{\sin\theta} - a E \sin \theta\right)^2  - X_0 a^2 \cos^2\theta =
-\Delta_r \left( \frac{df_r}{dr}\right)^2 + \frac{\Xi^2}{\Delta_r} [E (r^2+a^2) - a L_z]^2 + X_0 r^2 
\eqqcolon \mK,
\ee
where we have denoted the separation constant by $\mK$.
The solutions are given by 
\be \label{HJsol} 
f_r=\pm\int dr\frac{\sqrt{\mathcal{R}}}{\Delta_r}, \quad 
f_\theta=\pm\int d\theta\frac{\sqrt{\Theta}}{\Delta_\theta} , \ee
where
\begin{align} \label{RT}
\begin{split}
\mathcal{R}&= \Xi^2 [E (r^2+a^2) - a L_z]^2 + \Delta_r(X_0 r^2 -\mK), \\
\Theta&= -\Xi^2\left(\frac{L_z}{\sin\theta} - a E \sin \theta\right)^2  +  \Delta_\theta(X_0 a^2 \cos^2\theta+\mK) . 
\end{split}
\end{align}
These solutions determine the scalar field profile for a general asymptotically (anti-)de Sitter black hole solution characterized by mass, charge, and angular momentum in GR.
Taking an appropriate limit, one can recover special cases such as those studied in \S\ref{ssec:sss} or the Kerr-de Sitter solution obtained for a class of shift-symmetric DHOST theories with $A_1=A_2=0$ in \cite{Charmousis:2019vnf}.

The differential equation~\eqref{HJeq} coincides with the Hamilton-Jacobi equation for a test particle in the Kerr--Newman--(anti-)de Sitter background with the correspondence~$\phi=\mathcal{S}$ and $X_0 = -m^2$, where $\mathcal{S}$ is the Hamilton-Jacobi potential and $m(>0)$ is the mass of the test particle.
From Hamilton-Jacobi point of view, since the four-momentum is $p_\mu=\pa_\mu \mathcal{S}=\pa_\mu\phi$, we see that $p_t=-E$, $p_\varphi=L_z$, and $g^{\mu\nu}p_\mu p_\nu=-m^2=X_0$ correspond to three conserved quantities, where $E$ and $L_z$ are energy and angular momentum of the test particle. 
Furthermore, the separation constant $\mK$ is related to the Carter's constant~\cite{Carter:1968rr}.

In parallel to \cite{Charmousis:2019vnf}, requiring the regularity of the scalar field derivative, we can further restrict the functional form of the scalar field.
To make $\pa_\theta\phi= 0$ at  
$\theta=0$ and $\pi$, we restrict $L_z=0$ and $\mK=-X_0a^2=m^2a^2$, for which Eq.~\eqref{RT} is simplified as
\begin{align} \label{RT2}
\begin{split}
\mathcal{R}&= m^2(r^2 +a^2) [\eta^2 (r^2+a^2) - \Delta_r ], \\
\Theta&= m^2 a^2 \sin^2 \theta ( \Delta_\theta -\eta^2),
\end{split}
\end{align}
where $\eta\coloneqq \Xi E/m$.
When $\Lambda<0$, the range of $\Delta_r$ is all positive real numbers, so $\mathcal{R}$ cannot remain nonnegative.
Namely, the Kerr--Newman--anti-de Sitter spacetime cannot accommodate 
nontrivial scalar hair.
Hence, in what follows, we focus on asymptotically de Sitter spacetimes with $\Lambda>0$.
When $a\ne 0$, requiring $\mathcal{R}\geq 0$ and $\Theta\geq 0$, the parameter $\eta$ is constrained as 
$\sqrt{ \frac{\Delta_r}{r^2+a^2} } \leq |\eta| \leq \sqrt{\Delta_\theta}$ for any $r$ of interest for the lower bound and any $0\leq \theta \leq \pi$ for the upper bound.\footnote{When $X_0\ge 0$, we have
    \be
    \Theta=-a^2\sin^2\theta\bra{X_0\Delta_\theta+\Xi^2E^2}<0 \nonumber
    \ee
for a generic $\theta$, provided that $a\ne 0$ and $E\ne 0$.
Hence, for $a\ne 0$ and $E\ne 0$, $X_0$ should be negative.
Also, using the analogy of the Hamilton-Jacobi equation for a test particle with $X_0=-m^2$, a positive $X_0$ corresponds to an imaginary mass, which may be related to this issue.
Note also that, as long as $X_0=-m^2\ne 0$ and $a\ne 0$, a static scalar field with $\eta=0$ yields either $\mathcal{R}<0$ or $\Theta<0$.
This means that the Kerr-Newman-de Sitter metric with $a\ne 0$ is always accompanied by time-dependent scalar hair unless $X_0=0$. 
The case with a static scalar field and $X_0=0$ ends up with a constant scalar profile.
}
Consequently, we obtain $\eta_c \leq |\eta| \leq 1$. 
For $|\eta|=\eta_c$ there exists a finite $r_0$ such that $\mathcal{R}(r_0)=0$,
whereas for the upper bound $\Theta=0$ for some $\theta=\theta_0$. 
Note also that, for the Kerr-Newman case with $\Lambda=0$, the conditions~$\mathcal{R}\ge 0$ and $\Theta\ge 0$ restrict $|\eta|$ to be unity, under which we have $\Theta=0$.
This means that the Kerr-Newman metric cannot support $\theta$-dependent scalar hair.

Using $\Theta$ in \eqref{RT2}, we can analytically integrate $f_\theta$ in \eqref{HJsol} to obtain\footnote{The analysis for the Kerr-de Sitter solution performed in \cite{Charmousis:2019vnf} v1 is a special case of the analysis in the present paper, and here we note two typos in Eq.~(19) in \cite{Charmousis:2019vnf} v1; the overall factor and the additional $\eta$ in the first logarithmic function in our solution~\eqref{solfth2} were omitted.
}
    \be
    f_\theta=\pm m\sqrt{\fr{3}{\Lambda}}\brb{\eta\, {\rm artanh}\bra{\sqrt{\fr{\Lambda a^2}{3(\Delta_\theta-\eta^2)}}\eta\cos\theta}-{\rm arsinh}\bra{\sqrt{\fr{\Lambda a^2}{3(1-\eta^2)}}\cos\theta}} + {\rm const}, \label{solfth2}
    \ee
or equivalently, 
\be \label{solfth} f_\theta = \pm m \sqrt{\frac{3}{\Lambda}} \left[ \eta \log\left( \frac{ \sqrt{\Delta_\theta - \eta^2} + \eta \cos \theta \sqrt{\Lambda a^2/3} }{ \sqrt{(1-\eta^2)\Delta_\theta } } \right) 
- \log\left( \frac{ \sqrt{\Delta_\theta - \eta^2} + \cos \theta \sqrt{\Lambda a^2/3} }{ \sqrt{1-\eta^2} } \right)  \right] + {\rm const}. \ee
Note that each branch of $f_\theta$ is antisymmetric with respect to $\theta=\pi/2$ up to the constant offset.

In general, the square roots in \eqref{HJsol} should be defined carefully when $\Theta$ (or $\mathcal{R}$) vanishes for a certain value of $\theta$~(or $r$, respectively)~\cite{Charmousis:2019vnf}.
For instance, let us consider the case where $|\eta|=1$.
In this case, we have
    \be
    \Theta=\fr{1}{3}m^2a^4\Lambda\sin^2\theta\cos^2\theta,
    \ee
whose positive root is not smooth at $\theta=\pi/2$ where $\cos\theta$ changes its sign.
Thus, it is natural to take
    \be
    \sqrt{\Theta}=ma^2\sqrt{\fr{\Lambda}{3}}\sin\theta\cos\theta,
    \ee
and as a result we obtain\footnote{
Note that $f_\theta$ given by this functional form is symmetric with respect to $\theta=\pi/2$ as opposed to the case of generic $\eta$, meaning that the angular variation of the scalar field is symmetric about the equator if and only if $|\eta|=1$.
When $|\eta|<1$, one could change the branch at $\theta=\pi/2$ to make $f_\theta$ symmetric, which however makes $\partial_\theta \phi$ discontinuous at $\theta=\pi/2$.
}
    \be
    f_\theta = \mp \frac{m}{2}\sqrt{\frac{3}{\Lambda}} \log \Delta_\theta + {\rm const.} \label{fth-lim1}
    \ee
The expression~\eqref{fth-lim1} can be reproduced from \eqref{solfth2} by taking the limit~$|\eta|\to 1$.

The same notion on the branch choice applies to $\sqrt{\mathcal{R}}$ in \eqref{HJsol}.
As we mentioned above, for $\eta=\eta_c$ there exists a finite $r_0$ such that $\mathcal{R}(r_0)=0$, where one can change the branch to make the scalar field  smooth. As discussed in detail in \S\ref{ssec:sss}, this manipulation makes the scalar field profile finite at both the black hole and cosmological event horizons for a specific observer, and infinitely differentiable between the horizons.
This is a natural generalization of the argument in \cite{Charmousis:2019vnf}.

Let us consider several limiting cases of the parameters of the metric.
First, at the leading order in the limit~$\Lambda a^2\to 0$, we have
\be \label{limLa} f_\theta = \mp m a \sqrt{1 - \frac{E^2}{m^2}} \cos\theta \, [1+\mO(\Lambda a^2)] + {\rm const}. \ee
As expected, $f_\theta\to 0$ for $a\to 0$, which is consistent with the static spherically symmetric case in \S\ref{ssec:sss}. 
Note also that, as mentioned earlier, the possible range of $E^2/m^2$ depends on $\Lambda$ and $a$.
In the limit~$\Lambda\to 0$, the range shrinks to a single value~$E^2/m^2=1$, which means $f_\theta=0$.

For $a=0$ or the no-rotating case, 
the metric~\eqref{KNAdS} reduces to the Reissner--Nordstr\"om--(anti-)de Sitter one. 
In this case, $\Theta=0$ identically holds and 
\be \mathcal{R}=E^2r^4+X_0 r^2 \left[r^2\left(1-\frac{1}{3}\Lambda r^2\right) - r_g r + r_e^2 \right] . \ee
Note that $X_0=-m^2>0$ makes $\mathcal{R}$ positive, which allows a solution with a spacelike scalar field derivative.
This is consistent with the solutions in the literature~\cite{Minamitsuji:2018vuw,Motohashi:2019sen,Charmousis:2019vnf},
while for perturbations around the stealth Schwarzschild solution with $X_0>0$ in the Horndeski theory, it is known that one of the even-parity modes is strongly coupled~\cite{Minamitsuji:2018vuw}.
Solutions with $X_0\leq 0$ are also allowed so long as $\mathcal{R}\geq 0$.
This limit is also consistent with the analysis in \S\ref{ssec:sss}.

Next, let us consider several limiting cases of the scalar field ansatz.
For a static scalar field with $E=0$ or $\eta=0$, Eq.~\eqref{RT2} reads
    \be
    \mathcal{R}=-m^2(r^2+a^2)\Delta_r,\quad
    \Theta=m^2a^2\sin^2\theta\Delta_\theta.
    \ee
For $a\ne 0$, the conditions~$\mathcal{R}\ge 0$ and $\Theta\ge 0$ are compatible only if $m^2=0$, which results in $\phi={\rm const}$.

Let us then consider the case with a $\theta$-independent scalar field.
If we adopt the scalar field profile~\eqref{KNdSansatz} with $f_\theta={\rm const}$ from the outset, 
the fact that the left-hand side of \eqref{HJeq} is constant puts a constraint on the parameter space.
Provided that $a\ne 0$ and $\Xi>0$, Eq.~\eqref{HJeq} is satisfied if and only if
    \be
    X_0=E=L_z=\mK=0\quad {\rm or}\quad
    \Lambda=X_0+E^2=L_z=\mK-a^2E^2=0.
    \ee
In the first case, Eq.~\eqref{RT} reads $\mathcal{R}=0$, meaning that $f_r={\rm const}$.
Thus, we have a constant scalar field profile.
On the other hand, in the second case, the cosmological constant vanishes and the metric reduces to the Kerr-Newman one. 
This is consistent with the aforementioned fact that the Kerr-Newman metric cannot accommodate $\theta$-dependent scalar hair.
We then have 
    \be
    \mathcal{R}=E^2(r^2+a^2)(r_gr-r_e^2),
    \ee
from which one obtains $f_r$ in terms of elliptic functions.
The perturbation analysis for the second case with $r_e= 0$~(i.e., a hairy Kerr black hole) was performed in \cite{Charmousis:2019fre}.

Finally, let us consider the case with an $r$-independent scalar field profile. 
In this case, we can still use \eqref{RT2}, and thus we require that $\mathcal{R}$ given by this equation identically vanishes.
Provided that $\Xi>0$, we have the following constraints on the parameters:
    \be
    X_0=E=0\quad {\rm or}\quad
    r_g=r_e=\Lambda=X_0+E^2=0.
    \ee
In the first case, one ends up with $\phi={\rm const}$.
The second set of conditions forces the metric to be the massless Kerr one, and the scalar field can have only the $t$-dependent part, i.e., $\phi=-Et+{\rm const}$.

So far, we have considered rotating black hole solutions with scalar hair in four dimensions.
In three dimensions, the Ba{\~n}ados-Teitelboim-Zanelli~(BTZ) spacetime~\cite{Banados:1992wn},
\be g_{\mu\nu}dx^\mu dx^\nu=- A(r)dt^2 + \frac{dr^2}{A(r)}  + r^2 \left( \frac{J}{2r^2} dt-d\varphi \right)^2 , \quad 
A(r)= -M+\frac{r^2}{l^2}+\frac{J^2}{4r^2} - \frac{Q^2}{2}\log \left( \frac{r}{r_0} \right), \ee
where $\mu,\nu=0,1,2$,
is known as a general black hole solution in GR characterized by mass~$M$, charge~$Q$, and angular momentum~$J$ in the presence of a negative cosmological constant~$\Lambda=-l^{-2}$.
Here, the parameter~$r_0(>0)$ has been introduced just to nondimensionalize the argument of the logarithmic function, and its magnitude is irrelevant as any change of $r_0$ can be absorbed into the mass parameter~$M$.
The scalar field profile is given by
\be \phi=-Et+L_z\varphi+f_r(r), \quad 
f_r=\pm\int dr\frac{\sqrt{(J L_z - 2 Er^2)^2 + 4 r^2 (X_0r^2-L_z^2) A}}{2r^2A}, \ee
which is a natural generalization of the result in \cite{Bravo-Gaete:2014haa} obtained for a hairy non-charged BTZ black hole with $X={\rm const}$ in a class of the Horndeski theory.

\subsection{Homogeneous isotropic spacetime}\label{ssec:flrw}

Let us consider the Friedmann-Lema\^itre-Robertson-Walker (FLRW) spacetime,
\be g_{\mu\nu}dx^\mu dx^\nu=-dt^2+a(t)^2 \left[ \frac{dr^2}{1-kr^2} + r^2 (d\theta^2+\sin^2\theta d\varphi^2) \right] , \ee
where in this subsection $a=a(t)$ denotes the scale factor and $k=0,\pm 1$ denotes the spatial curvature.

We can obtain the scalar field profile with a certain ansatz.
Adopting an $r$-independent scalar field profile~$\phi=\phi(t)$ and assuming $X=X_0\le 0$, we simply obtain 
\be \label{FLRWsol1} \phi = \pm \sqrt{-X_0} \, t , \ee
regardless of the spatial curvature or the evolution of the scale factor.

A different ansatz leads to a different class of solutions.
As a demonstration, let us adopt the following scalar field profile:
\be \phi=\xi(t)+\psi(r). \ee
The equation~$X=X_0$ is then separable as 
\be (1 - k r^2) \psi'{}^2 = a^2 (X_0 + \dot\xi^2) \eqqcolon \mC^2 , \ee
where $\mC(>0)$ is the separation constant.  
Note that we can take the static limit by sending $\mC\to0$ and recover \eqref{FLRWsol1}. 
We then obtain 
\be \label{FLRWsol2} \psi(r)=\pm \mC \frac{ {\rm arcsin} (\sqrt{k} r)}{\sqrt{k}} , \quad
\xi(t)= \pm \int dt \sqrt{\frac{\mC^2}{a^2(t)}-X_0} . 
\ee
The solution for $\psi(r)$ should be understood as $\psi(r)= \pm \mC r,\pm \mC {\rm arcsin}\, r,\pm \mC {\rm arsinh}\, r$ for $k=0,+1,-1$, respectively.
Here, we assume $X_0\le 0$ so that $\xi$ is always real.
The evolution of $\xi(t)$ depends on the evolution of the scale factor~$a(t)$, which is determined by the Friedmann equation with the spatial curvature and a matter component.
One can perform the integral analytically to obtain $\xi(t)$ for several simple cases such as the de Sitter expansion~$a(t)=e^{Ht}$,
\be \xi(t)= \mp H^{-1} \left[ \sqrt{\mC^2 e^{-2Ht} -X_0} -\sqrt{-X_0}\,{\rm arsinh} \bra{\mC^{-1}\sqrt{-X_0} e^{Ht} } \right]+{\rm const}, \ee
or the power-law expansion~$a(t)=a_0(t/t_0)^p$ with $p>0$,
    \be
    \xi(t)=\left\{
    \begin{array}{ll}
    \displaystyle \pm \fr{\mC t_0}{(1-p)a_0}\bra{\fr{t}{t_0}}^{1-p}
    {}_2F_1\bra{-\fr{1}{2},-\fr{1}{2}+\fr{1}{2p};\fr{1}{2}+\fr{1}{2p};\fr{a_0^2X_0}{\mC^2}\bra{\fr{t}{t_0}}^{2p}}+{\rm const}
    &{\rm for}~p\ne 1, \\
    \displaystyle \pm \fr{\mC t_0}{a_0}\brb{\sqrt{1-\fr{a_0^2X_0}{\mC^2}\bra{\fr{t}{t_0}}^{2}}-{\rm artanh}\sqrt{1-\fr{a_0^2X_0}{\mC^2}\bra{\fr{t}{t_0}}^{2}}}+{\rm const}
    &{\rm for}~p=1, \\
    \end{array}\right. \label{power-law2}
    \ee
where ${}_2F_1$ denotes the Gauss' hypergeometric function.
For a general expansion history, one needs to perform the integral numerically.

The closed FLRW spacetime with $k=+1$ can be used to describe a collapsing object.
Indeed, the Oppenheimer-Snyder collapse~\cite{Oppenheimer:1939ue} and its generalization to the case with a cosmological constant~\cite{Nakao:1991sh}
are known as the standard approach to describe the spherical collapse,
and they consist of the closed FLRW spacetime for the interior of a dust sphere and the Schwarzschild(-de Sitter) spacetime for the exterior region.
Given that both the metrics are a solution of a given scalar-tensor theory, such collapsing solutions are also allowed with the scalar field profile obtained above for each region.

\section{Conclusions}\label{sec:con}

In this paper, we have explored exact solutions of the same form as in GR in general quadratic higher-order scalar-tensor theories.
Adopting the ansatz of constant kinetic term, we have derived a set of conditions in a fully covariant manner, under which the EL equations allow 
GR solutions as exact solutions in the presence of a general matter component, generalizing the covariant analysis in \cite{Motohashi:2018wdq} with the ansatz of constant scalar field.
Adopting an appropriate ansatz, the scalar field profile can be obtained by integrating the condition of constant kinetic term for each metric solution.
We have demonstrated the construction of the scalar field profile for several cases including the 
Kerr-Newman-de Sitter spacetime, i.e., black hole solutions characterized by mass, charge, and angular momentum in the presence of a cosmological constant.
For the marginal values of the parameters, one should choose the branch of the radial part of $\phi$ appropriately, or otherwise the scalar field is not smooth at some finite $r$ or $\theta$. 
Also, a careful analysis is needed when one starts from some general setup and then restricts oneself to a limited case with a stronger ansatz for the metric or scalar field, since the metric ansatz may in general constrain the scalar field profile, and vice versa.
Furthermore, we showed that asymptotically anti-de Sitter spacetimes cannot support 
nontrivial scalar hair. 
Our condition for the existence of GR solutions also applies to more general solutions such as the Pleba\'nski-Demia\'nski solution~\cite{Plebanski:1976gy}.
Furthermore, it applies to not only black hole solutions but also any solution in GR in the presence of a matter component.

The stability of the exact solutions obtained in the present paper would be one of the most important issues that need to be clarified.
In general, even though the background spacetime is 
that of GR solutions, perturbations can behave differently.
It would thus be intriguing to investigate the evolution of perturbations around the exact solutions 
and clarify how the difference can show up.
For some class of stealth solutions, the scalar field perturbation is strongly coupled~\cite{Minamitsuji:2018vuw,deRham:2019slh}, but for the case with a timelike scalar field derivative, the problem of strong coupling can be avoided by the scordatura mechanism~\cite{Motohashi:2019ymr}, which may also cause a distinctive signature.
Also, the strong coupling issue would be absent in theories where the scalar degree of freedom does not propagate~\cite{Lin:2017oow,Chagoya:2018yna,Aoki:2018zcv,Afshordi:2006ad,Iyonaga:2018vnu,Gao:2019twq}.
Another thing of interest is to exploit the exact solutions as a seed to generate a new solution using the conformal/disformal transformation~\cite{BenAchour:2019fdf} or the Kerr-Schild transformation~\cite{Babichev:2020qpr}.
For instance, one can generate a deformed Kerr solution and see how the physics changes~\cite{BenAchour:2020fgy}.
Our approach can also be extended to theories involving higher-order interactions of second derivatives or even higher derivatives of $\phi$.
We leave these issues as a future work.

\acknowledgments
We thank Christos Charmousis and Marco Crisostomi for useful comments.
K.T.\ was supported by Japan Society for the Promotion of Science (JSPS) Grants-in-Aid for Scientific Research (KAKENHI) No.\ JP17H02894 and No.\ JP17K18778.
H.M.\ was supported by JSPS KAKENHI Grant No.\ JP17H06359 and No.\ JP18K13565.

\appendix*

\section{Existence Conditions for General Relativity solutions in arbitrary dimensions}
\label{App}

In \S\ref{sec:cov}, we derived the set of conditions~\eqref{exist_cond_HOST} for quadratic higher-derivative scalar-tensor theories described by the action~\eqref{qHOST} to accommodate GR solutions in four spacetime dimensions.
Interestingly, one can generalize the discussion to arbitrary $D$ dimensions, with $D\ge 3$.
Indeed, with the ansatz~$X=X_0={\rm const}$, one obtains the same EL equation for the metric as \eqref{EmnHOST}, and the only modification arises from the following equations:
    \be
    R_\mn=8\pi G\bra{T_\mn-\fr{1}{D-2}Tg_\mn}+\fr{2}{D-2}\Lambda g_\mn,\quad
    R=-\fr{2}{D-2}\bra{8\pi GT-D\Lambda},
    \ee
which originate from the Einstein equation in GR, $G_{\mu\nu} = 8\pi G T_{\mu\nu} - \Lambda g_{\mu\nu}$, with $G$ being the gravitational constant in $D$ dimensions.
Then, the existence conditions for GR solutions in $D$ dimensions are obtained as follows:
\begin{align} 
\begin{split}
&F_0+2\Lambda F_2 - X_0 (F_{1\phi} + 2 F_{2\phi\phi}) = 0,\quad
2F_{1X} + 4A_{1\phi} - X_0A_{3\phi}=0, \quad
F_{2\phi} - X_0 A_{1\phi} =0 , \\
&A_1 = 0, \quad
A_2 = 0, \quad
A_{1\phi}+A_{2\phi} = 0, \quad
A_{1X}+A_{2X} = 0, \quad
A_3+2A_{1X} = 0 , \quad
\bra{16\pi GF_2 - 1} T_{\mu\nu} = 0, \\
&F_{0X}-F_{1\phi} - F_{2\phi\phi} + \fr{2\Lambda}{D-2} \bra{D F_{2X} - X_0 A_{1X}} = 8\pi G\brb{ \fr{2}{D-2}F_{2X}T + A_{1X} \bra{\phi^\alpha\phi^\beta T_{\alpha\beta}-\fr{1}{D-2}TX_0}}, \label{exist_cond_HOST_D}
\end{split}
\end{align}
where the coupling functions and their derivatives should be evaluated at $X=X_0$.
One can check that this set of conditions reduces to \eqref{exist_cond_HOST} when $D=4$.
It is also straightforward to generalize the conditions~\eqref{exist_cond_phi_const} for the constant scalar profile~$\phi=\phi_0$ to $D$ dimensions, which are summarized as
\begin{align}
    F_0+2\Lambda F_2 = 0 ,\quad
    F_{0\phi}+\fr{2D\Lambda}{D-2} F_{2\phi}= 0,\quad
    \bra{16\pi GF_2-1}T_\mn=0,\quad
    F_{2\phi} T =0, \label{exist_cond_phi_const_D}
\end{align}
where the coupling functions are evaluated at $(\phi,X)=(\phi_0,0)$.

Note in passing that the case of $D=2$ should be treated separately as the Einstein-Hilbert action becomes a total derivative, and one needs to consider an alternative theory.
If one chooses a two-dimensional action and EL equations analogous to GR, one can derive conditions in parallel to \eqref{exist_cond_HOST_D} or \eqref{exist_cond_phi_const_D} under which the higher-order scalar-tensor theories allow exact solutions of the same form as those in the two-dimensional theory.
For general black hole solutions with $X={\rm const}$ in general two-dimensional scalar-tensor theories, see \cite{Takahashi:2018yzc}.


\bibliographystyle{mybibstyle}
\bibliography{rotatingBH}

\begin{thebibliography}{74}%
\makeatletter
\providecommand \@ifxundefined [1]{%
 \@ifx{#1\undefined}
}%
\providecommand \@ifnum [1]{%
 \ifnum #1\expandafter \@firstoftwo
 \else \expandafter \@secondoftwo
 \fi
}%
\providecommand \@ifx [1]{%
 \ifx #1\expandafter \@firstoftwo
 \else \expandafter \@secondoftwo
 \fi
}%
\providecommand \natexlab [1]{#1}%
\providecommand \enquote  [1]{``#1''}%
\providecommand \bibnamefont  [1]{#1}%
\providecommand \bibfnamefont [1]{#1}%
\providecommand \citenamefont [1]{#1}%
\providecommand \href@noop [0]{\@secondoftwo}%
\providecommand \href [0]{\begingroup \@sanitize@url \@href}%
\providecommand \@href[1]{\@@startlink{#1}\@@href}%
\providecommand \@@href[1]{\endgroup#1\@@endlink}%
\providecommand \@sanitize@url [0]{\catcode `\\12\catcode `\$12\catcode
  `\&12\catcode `\#12\catcode `\^12\catcode `\_12\catcode `\%12\relax}%
\providecommand \@@startlink[1]{}%
\providecommand \@@endlink[0]{}%
\providecommand \url  [0]{\begingroup\@sanitize@url \@url }%
\providecommand \@url [1]{\endgroup\@href {#1}{\urlprefix }}%
\providecommand \urlprefix  [0]{URL }%
\providecommand \Eprint [0]{\href }%
\providecommand \doibase [0]{http://dx.doi.org/}%
\providecommand \selectlanguage [0]{\@gobble}%
\providecommand \bibinfo  [0]{\@secondoftwo}%
\providecommand \bibfield  [0]{\@secondoftwo}%
\providecommand \translation [1]{[#1]}%
\providecommand \BibitemOpen [0]{}%
\providecommand \bibitemStop [0]{}%
\providecommand \bibitemNoStop [0]{.\EOS\space}%
\providecommand \EOS [0]{\spacefactor3000\relax}%
\providecommand \BibitemShut  [1]{\csname bibitem#1\endcsname}%
\let\auto@bib@innerbib\@empty
\bibitem [{\citenamefont {Abbott}\ \emph {et~al.}(2016)\citenamefont {Abbott}
  \emph {et~al.}}]{Abbott:2016blz}%
  \BibitemOpen
  \bibfield  {author} {\bibinfo {author} {\bibfnamefont {B.~P.}\ \bibnamefont
  {Abbott}} \emph {et~al.} (\bibinfo {collaboration} {LIGO Scientific,
  Virgo}),\ }\href {\doibase 10.1103/PhysRevLett.116.061102} {\bibfield
  {journal} {\bibinfo  {journal} {\emph {Phys. Rev. Lett.}}\ }\textbf {\bibinfo
  {volume} {116}},\ \bibinfo {pages} {061102} (\bibinfo {year} {2016})},\
  \Eprint {http://arxiv.org/abs/1602.03837} {arXiv:1602.03837 [gr-qc]}
  \BibitemShut {NoStop}%
\bibitem [{\citenamefont {Stephani}\ \emph {et~al.}(2003)\citenamefont
  {Stephani}, \citenamefont {Kramer}, \citenamefont {MacCallum}, \citenamefont
  {Hoenselaers},\ and\ \citenamefont {Herlt}}]{Stephani:2003tm}%
  \BibitemOpen
  \bibfield  {author} {\bibinfo {author} {\bibfnamefont {H.}~\bibnamefont
  {Stephani}}, \bibinfo {author} {\bibfnamefont {D.}~\bibnamefont {Kramer}},
  \bibinfo {author} {\bibfnamefont {M.~A.~H.}\ \bibnamefont {MacCallum}},
  \bibinfo {author} {\bibfnamefont {C.}~\bibnamefont {Hoenselaers}},  and
  \bibinfo {author} {\bibfnamefont {E.}~\bibnamefont {Herlt}},\ }\href
  {\doibase 10.1017/CBO9780511535185} {\emph {\bibinfo {title} {{Exact
  solutions of Einstein's field equations}}}},\ Cambridge Monographs on
  Mathematical Physics\ (\bibinfo  {publisher} {Cambridge Univ. Press},\
  \bibinfo {address} {Cambridge},\ \bibinfo {year} {2003})\BibitemShut
  {NoStop}%
\bibitem [{\citenamefont {Griffiths}\ and\ \citenamefont
  {Podolsk{\'y}}(2009)}]{Griffiths:2009dfa}%
  \BibitemOpen
  \bibfield  {author} {\bibinfo {author} {\bibfnamefont {J.~B.}\ \bibnamefont
  {Griffiths}} and \bibinfo {author} {\bibfnamefont {J.}~\bibnamefont
  {Podolsk{\'y}}},\ }\href {\doibase 10.1017/CBO9780511635397} {\emph {\bibinfo
  {title} {{Exact Space-Times in Einstein's General Relativity}}}},\ Cambridge
  Monographs on Mathematical Physics\ (\bibinfo  {publisher} {Cambridge
  University Press},\ \bibinfo {address} {Cambridge},\ \bibinfo {year}
  {2009})\BibitemShut {NoStop}%
\bibitem [{\citenamefont {Motohashi}\ and\ \citenamefont
  {Minamitsuji}(2018)}]{Motohashi:2018wdq}%
  \BibitemOpen
  \bibfield  {author} {\bibinfo {author} {\bibfnamefont {H.}~\bibnamefont
  {Motohashi}} and \bibinfo {author} {\bibfnamefont {M.}~\bibnamefont
  {Minamitsuji}},\ }\href {\doibase 10.1016/j.physletb.2018.04.041} {\bibfield
  {journal} {\bibinfo  {journal} {\emph {Phys. Lett. B}}\ }\textbf {\bibinfo
  {volume} {781}},\ \bibinfo {pages} {728} (\bibinfo {year} {2018})},\ \Eprint
  {http://arxiv.org/abs/1804.01731} {arXiv:1804.01731 [gr-qc]} \BibitemShut
  {NoStop}%
\bibitem [{\citenamefont {Woodard}(2015)}]{Woodard:2015zca}%
  \BibitemOpen
  \bibfield  {author} {\bibinfo {author} {\bibfnamefont {R.~P.}\ \bibnamefont
  {Woodard}},\ }\href {\doibase 10.4249/scholarpedia.32243} {\bibfield
  {journal} {\bibinfo  {journal} {\emph {Scholarpedia}}\ }\textbf {\bibinfo
  {volume} {10}},\ \bibinfo {pages} {32243} (\bibinfo {year} {2015})},\ \Eprint
  {http://arxiv.org/abs/1506.02210} {arXiv:1506.02210 [hep-th]} \BibitemShut
  {NoStop}%
\bibitem [{\citenamefont {Raidal}\ and\ \citenamefont
  {Veerm{\"a}e}(2017)}]{Raidal:2016wop}%
  \BibitemOpen
  \bibfield  {author} {\bibinfo {author} {\bibfnamefont {M.}~\bibnamefont
  {Raidal}} and \bibinfo {author} {\bibfnamefont {H.}~\bibnamefont
  {Veerm{\"a}e}},\ }\href {\doibase 10.1016/j.nuclphysb.2017.01.024} {\bibfield
   {journal} {\bibinfo  {journal} {\emph {Nucl. Phys. B}}\ }\textbf {\bibinfo
  {volume} {916}},\ \bibinfo {pages} {607} (\bibinfo {year} {2017})},\ \Eprint
  {http://arxiv.org/abs/1611.03498} {arXiv:1611.03498 [hep-th]} \BibitemShut
  {NoStop}%
\bibitem [{\citenamefont {Smilga}(2017)}]{Smilga:2017arl}%
  \BibitemOpen
  \bibfield  {author} {\bibinfo {author} {\bibfnamefont {A.}~\bibnamefont
  {Smilga}},\ }\href {\doibase 10.1142/S0217751X17300253} {\bibfield  {journal}
  {\bibinfo  {journal} {\emph {Int. J. Mod. Phys. A}}\ }\textbf {\bibinfo
  {volume} {32}},\ \bibinfo {pages} {1730025} (\bibinfo {year} {2017})},\
  \Eprint {http://arxiv.org/abs/1710.11538} {arXiv:1710.11538 [hep-th]}
  \BibitemShut {NoStop}%
\bibitem [{\citenamefont {Motohashi}\ and\ \citenamefont
  {Suyama}(2020)}]{Motohashi:2020psc}%
  \BibitemOpen
  \bibfield  {author} {\bibinfo {author} {\bibfnamefont {H.}~\bibnamefont
  {Motohashi}} and \bibinfo {author} {\bibfnamefont {T.}~\bibnamefont
  {Suyama}},\ }\Eprint {http://arxiv.org/abs/2001.02483} {arXiv:2001.02483
  [hep-th]} \BibitemShut {NoStop}%
\bibitem [{\citenamefont {Motohashi}\ and\ \citenamefont
  {Suyama}(2015)}]{Motohashi:2014opa}%
  \BibitemOpen
  \bibfield  {author} {\bibinfo {author} {\bibfnamefont {H.}~\bibnamefont
  {Motohashi}} and \bibinfo {author} {\bibfnamefont {T.}~\bibnamefont
  {Suyama}},\ }\href {\doibase 10.1103/PhysRevD.91.085009} {\bibfield
  {journal} {\bibinfo  {journal} {\emph {Phys. Rev. D}}\ }\textbf {\bibinfo
  {volume} {91}},\ \bibinfo {pages} {085009} (\bibinfo {year} {2015})},\
  \Eprint {http://arxiv.org/abs/1411.3721} {arXiv:1411.3721 [physics.class-ph]}
  \BibitemShut {NoStop}%
\bibitem [{\citenamefont {Langlois}\ and\ \citenamefont
  {Noui}(2016)}]{Langlois:2015cwa}%
  \BibitemOpen
  \bibfield  {author} {\bibinfo {author} {\bibfnamefont {D.}~\bibnamefont
  {Langlois}} and \bibinfo {author} {\bibfnamefont {K.}~\bibnamefont {Noui}},\
  }\href {\doibase 10.1088/1475-7516/2016/02/034} {\bibfield  {journal}
  {\bibinfo  {journal} {\emph {JCAP}}\ }\textbf {\bibinfo {volume} {02}},\
  \bibinfo {pages} {034} (\bibinfo {year} {2016})},\ \Eprint
  {http://arxiv.org/abs/1510.06930} {arXiv:1510.06930 [gr-qc]} \BibitemShut
  {NoStop}%
\bibitem [{\citenamefont {Motohashi}\ \emph
  {et~al.}(2016{\natexlab{a}})\citenamefont {Motohashi}, \citenamefont {Noui},
  \citenamefont {Suyama}, \citenamefont {Yamaguchi},\ and\ \citenamefont
  {Langlois}}]{Motohashi:2016ftl}%
  \BibitemOpen
  \bibfield  {author} {\bibinfo {author} {\bibfnamefont {H.}~\bibnamefont
  {Motohashi}}, \bibinfo {author} {\bibfnamefont {K.}~\bibnamefont {Noui}},
  \bibinfo {author} {\bibfnamefont {T.}~\bibnamefont {Suyama}}, \bibinfo
  {author} {\bibfnamefont {M.}~\bibnamefont {Yamaguchi}},  and \bibinfo
  {author} {\bibfnamefont {D.}~\bibnamefont {Langlois}},\ }\href {\doibase
  10.1088/1475-7516/2016/07/033} {\bibfield  {journal} {\bibinfo  {journal}
  {\emph {JCAP}}\ }\textbf {\bibinfo {volume} {07}},\ \bibinfo {pages} {033}
  (\bibinfo {year} {2016}{\natexlab{a}})},\ \Eprint
  {http://arxiv.org/abs/1603.09355} {arXiv:1603.09355 [hep-th]} \BibitemShut
  {NoStop}%
\bibitem [{\citenamefont {Klein}\ and\ \citenamefont
  {Roest}(2016)}]{Klein:2016aiq}%
  \BibitemOpen
  \bibfield  {author} {\bibinfo {author} {\bibfnamefont {R.}~\bibnamefont
  {Klein}} and \bibinfo {author} {\bibfnamefont {D.}~\bibnamefont {Roest}},\
  }\href {\doibase 10.1007/JHEP07(2016)130} {\bibfield  {journal} {\bibinfo
  {journal} {\emph {JHEP}}\ }\textbf {\bibinfo {volume} {07}},\ \bibinfo
  {pages} {130} (\bibinfo {year} {2016})},\ \Eprint
  {http://arxiv.org/abs/1604.01719} {arXiv:1604.01719 [hep-th]} \BibitemShut
  {NoStop}%
\bibitem [{\citenamefont {Motohashi}\ \emph
  {et~al.}(2018{\natexlab{a}})\citenamefont {Motohashi}, \citenamefont
  {Suyama},\ and\ \citenamefont {Yamaguchi}}]{Motohashi:2017eya}%
  \BibitemOpen
  \bibfield  {author} {\bibinfo {author} {\bibfnamefont {H.}~\bibnamefont
  {Motohashi}}, \bibinfo {author} {\bibfnamefont {T.}~\bibnamefont {Suyama}},
  and \bibinfo {author} {\bibfnamefont {M.}~\bibnamefont {Yamaguchi}},\ }\href
  {\doibase 10.7566/JPSJ.87.063401} {\bibfield  {journal} {\bibinfo  {journal}
  {\emph {J. Phys. Soc. Jap.}}\ }\textbf {\bibinfo {volume} {87}},\ \bibinfo
  {pages} {063401} (\bibinfo {year} {2018}{\natexlab{a}})},\ \Eprint
  {http://arxiv.org/abs/1711.08125} {arXiv:1711.08125 [hep-th]} \BibitemShut
  {NoStop}%
\bibitem [{\citenamefont {Motohashi}\ \emph
  {et~al.}(2018{\natexlab{b}})\citenamefont {Motohashi}, \citenamefont
  {Suyama},\ and\ \citenamefont {Yamaguchi}}]{Motohashi:2018pxg}%
  \BibitemOpen
  \bibfield  {author} {\bibinfo {author} {\bibfnamefont {H.}~\bibnamefont
  {Motohashi}}, \bibinfo {author} {\bibfnamefont {T.}~\bibnamefont {Suyama}},
  and \bibinfo {author} {\bibfnamefont {M.}~\bibnamefont {Yamaguchi}},\ }\href
  {\doibase 10.1007/JHEP06(2018)133} {\bibfield  {journal} {\bibinfo  {journal}
  {\emph {JHEP}}\ }\textbf {\bibinfo {volume} {06}},\ \bibinfo {pages} {133}
  (\bibinfo {year} {2018}{\natexlab{b}})},\ \Eprint
  {http://arxiv.org/abs/1804.07990} {arXiv:1804.07990 [hep-th]} \BibitemShut
  {NoStop}%
\bibitem [{\citenamefont {Crisostomi}\ \emph {et~al.}(2016)\citenamefont
  {Crisostomi}, \citenamefont {Koyama},\ and\ \citenamefont
  {Tasinato}}]{Crisostomi:2016czh}%
  \BibitemOpen
  \bibfield  {author} {\bibinfo {author} {\bibfnamefont {M.}~\bibnamefont
  {Crisostomi}}, \bibinfo {author} {\bibfnamefont {K.}~\bibnamefont {Koyama}},
  and \bibinfo {author} {\bibfnamefont {G.}~\bibnamefont {Tasinato}},\ }\href
  {\doibase 10.1088/1475-7516/2016/04/044} {\bibfield  {journal} {\bibinfo
  {journal} {\emph {JCAP}}\ }\textbf {\bibinfo {volume} {04}},\ \bibinfo
  {pages} {044} (\bibinfo {year} {2016})},\ \Eprint
  {http://arxiv.org/abs/1602.03119} {arXiv:1602.03119 [hep-th]} \BibitemShut
  {NoStop}%
\bibitem [{\citenamefont {Ben~Achour}\ \emph
  {et~al.}(2016{\natexlab{a}})\citenamefont {Ben~Achour}, \citenamefont
  {Crisostomi}, \citenamefont {Koyama}, \citenamefont {Langlois}, \citenamefont
  {Noui},\ and\ \citenamefont {Tasinato}}]{BenAchour:2016fzp}%
  \BibitemOpen
  \bibfield  {author} {\bibinfo {author} {\bibfnamefont {J.}~\bibnamefont
  {Ben~Achour}}, \bibinfo {author} {\bibfnamefont {M.}~\bibnamefont
  {Crisostomi}}, \bibinfo {author} {\bibfnamefont {K.}~\bibnamefont {Koyama}},
  \bibinfo {author} {\bibfnamefont {D.}~\bibnamefont {Langlois}}, \bibinfo
  {author} {\bibfnamefont {K.}~\bibnamefont {Noui}},  and \bibinfo {author}
  {\bibfnamefont {G.}~\bibnamefont {Tasinato}},\ }\href {\doibase
  10.1007/JHEP12(2016)100} {\bibfield  {journal} {\bibinfo  {journal} {\emph
  {JHEP}}\ }\textbf {\bibinfo {volume} {12}},\ \bibinfo {pages} {100} (\bibinfo
  {year} {2016}{\natexlab{a}})},\ \Eprint {http://arxiv.org/abs/1608.08135}
  {arXiv:1608.08135 [hep-th]} \BibitemShut {NoStop}%
\bibitem [{\citenamefont {Takahashi}\ and\ \citenamefont
  {Kobayashi}(2017)}]{Takahashi:2017pje}%
  \BibitemOpen
  \bibfield  {author} {\bibinfo {author} {\bibfnamefont {K.}~\bibnamefont
  {Takahashi}} and \bibinfo {author} {\bibfnamefont {T.}~\bibnamefont
  {Kobayashi}},\ }\href {\doibase 10.1088/1475-7516/2017/11/038} {\bibfield
  {journal} {\bibinfo  {journal} {\emph {JCAP}}\ }\textbf {\bibinfo {volume}
  {11}},\ \bibinfo {pages} {038} (\bibinfo {year} {2017})},\ \Eprint
  {http://arxiv.org/abs/1708.02951} {arXiv:1708.02951 [gr-qc]} \BibitemShut
  {NoStop}%
\bibitem [{\citenamefont {Langlois}\ \emph {et~al.}(2019)\citenamefont
  {Langlois}, \citenamefont {Mancarella}, \citenamefont {Noui},\ and\
  \citenamefont {Vernizzi}}]{Langlois:2018jdg}%
  \BibitemOpen
  \bibfield  {author} {\bibinfo {author} {\bibfnamefont {D.}~\bibnamefont
  {Langlois}}, \bibinfo {author} {\bibfnamefont {M.}~\bibnamefont
  {Mancarella}}, \bibinfo {author} {\bibfnamefont {K.}~\bibnamefont {Noui}},
  and \bibinfo {author} {\bibfnamefont {F.}~\bibnamefont {Vernizzi}},\ }\href
  {\doibase 10.1088/1475-7516/2019/02/036} {\bibfield  {journal} {\bibinfo
  {journal} {\emph {JCAP}}\ }\textbf {\bibinfo {volume} {02}},\ \bibinfo
  {pages} {036} (\bibinfo {year} {2019})},\ \Eprint
  {http://arxiv.org/abs/1802.03394} {arXiv:1802.03394 [gr-qc]} \BibitemShut
  {NoStop}%
\bibitem [{\citenamefont {Gao}(2014)}]{Gao:2014soa}%
  \BibitemOpen
  \bibfield  {author} {\bibinfo {author} {\bibfnamefont {X.}~\bibnamefont
  {Gao}},\ }\href {\doibase 10.1103/PhysRevD.90.081501} {\bibfield  {journal}
  {\bibinfo  {journal} {\emph {Phys. Rev. D}}\ }\textbf {\bibinfo {volume}
  {90}},\ \bibinfo {pages} {081501} (\bibinfo {year} {2014})},\ \Eprint
  {http://arxiv.org/abs/1406.0822} {arXiv:1406.0822 [gr-qc]} \BibitemShut
  {NoStop}%
\bibitem [{\citenamefont {De~Felice}\ \emph {et~al.}(2018)\citenamefont
  {De~Felice}, \citenamefont {Langlois}, \citenamefont {Mukohyama},
  \citenamefont {Noui},\ and\ \citenamefont {Wang}}]{DeFelice:2018mkq}%
  \BibitemOpen
  \bibfield  {author} {\bibinfo {author} {\bibfnamefont {A.}~\bibnamefont
  {De~Felice}}, \bibinfo {author} {\bibfnamefont {D.}~\bibnamefont {Langlois}},
  \bibinfo {author} {\bibfnamefont {S.}~\bibnamefont {Mukohyama}}, \bibinfo
  {author} {\bibfnamefont {K.}~\bibnamefont {Noui}},  and \bibinfo {author}
  {\bibfnamefont {A.}~\bibnamefont {Wang}},\ }\href {\doibase
  10.1103/PhysRevD.98.084024} {\bibfield  {journal} {\bibinfo  {journal} {\emph
  {Phys. Rev. D}}\ }\textbf {\bibinfo {volume} {98}},\ \bibinfo {pages}
  {084024} (\bibinfo {year} {2018})},\ \Eprint
  {http://arxiv.org/abs/1803.06241} {arXiv:1803.06241 [hep-th]} \BibitemShut
  {NoStop}%
\bibitem [{\citenamefont {Gao}\ and\ \citenamefont {Yao}(2019)}]{Gao:2018znj}%
  \BibitemOpen
  \bibfield  {author} {\bibinfo {author} {\bibfnamefont {X.}~\bibnamefont
  {Gao}} and \bibinfo {author} {\bibfnamefont {Z.-B.}\ \bibnamefont {Yao}},\
  }\href {\doibase 10.1088/1475-7516/2019/05/024} {\bibfield  {journal}
  {\bibinfo  {journal} {\emph {JCAP}}\ }\textbf {\bibinfo {volume} {05}},\
  \bibinfo {pages} {024} (\bibinfo {year} {2019})},\ \Eprint
  {http://arxiv.org/abs/1806.02811} {arXiv:1806.02811 [gr-qc]} \BibitemShut
  {NoStop}%
\bibitem [{\citenamefont {Motohashi}\ and\ \citenamefont
  {Hu}(2020)}]{Motohashi:2020wxj}%
  \BibitemOpen
  \bibfield  {author} {\bibinfo {author} {\bibfnamefont {H.}~\bibnamefont
  {Motohashi}} and \bibinfo {author} {\bibfnamefont {W.}~\bibnamefont {Hu}},\
  }\href {\doibase 10.1103/PhysRevD.101.083531} {\bibfield  {journal} {\bibinfo
   {journal} {\emph {Phys. Rev. D}}\ }\textbf {\bibinfo {volume} {101}},\
  \bibinfo {pages} {083531} (\bibinfo {year} {2020})},\ \Eprint
  {http://arxiv.org/abs/2002.07967} {arXiv:2002.07967 [hep-th]} \BibitemShut
  {NoStop}%
\bibitem [{\citenamefont {Horndeski}(1974)}]{Horndeski:1974wa}%
  \BibitemOpen
  \bibfield  {author} {\bibinfo {author} {\bibfnamefont {G.~W.}\ \bibnamefont
  {Horndeski}},\ }\href {\doibase 10.1007/BF01807638} {\bibfield  {journal}
  {\bibinfo  {journal} {\emph {Int. J. Theor. Phys.}}\ }\textbf {\bibinfo
  {volume} {10}},\ \bibinfo {pages} {363} (\bibinfo {year} {1974})}\BibitemShut
  {NoStop}%
\bibitem [{\citenamefont {Nicolis}\ \emph {et~al.}(2009)\citenamefont
  {Nicolis}, \citenamefont {Rattazzi},\ and\ \citenamefont
  {Trincherini}}]{Nicolis:2008in}%
  \BibitemOpen
  \bibfield  {author} {\bibinfo {author} {\bibfnamefont {A.}~\bibnamefont
  {Nicolis}}, \bibinfo {author} {\bibfnamefont {R.}~\bibnamefont {Rattazzi}},
  and \bibinfo {author} {\bibfnamefont {E.}~\bibnamefont {Trincherini}},\
  }\href {\doibase 10.1103/PhysRevD.79.064036} {\bibfield  {journal} {\bibinfo
  {journal} {\emph {Phys. Rev. D}}\ }\textbf {\bibinfo {volume} {79}},\
  \bibinfo {pages} {064036} (\bibinfo {year} {2009})},\ \Eprint
  {http://arxiv.org/abs/0811.2197} {arXiv:0811.2197 [hep-th]} \BibitemShut
  {NoStop}%
\bibitem [{\citenamefont {Deffayet}\ \emph
  {et~al.}(2009{\natexlab{a}})\citenamefont {Deffayet}, \citenamefont
  {Esposito-Far{\`e}se},\ and\ \citenamefont {Vikman}}]{Deffayet:2009wt}%
  \BibitemOpen
  \bibfield  {author} {\bibinfo {author} {\bibfnamefont {C.}~\bibnamefont
  {Deffayet}}, \bibinfo {author} {\bibfnamefont {G.}~\bibnamefont
  {Esposito-Far{\`e}se}},  and \bibinfo {author} {\bibfnamefont
  {A.}~\bibnamefont {Vikman}},\ }\href {\doibase 10.1103/PhysRevD.79.084003}
  {\bibfield  {journal} {\bibinfo  {journal} {\emph {Phys. Rev. D}}\ }\textbf
  {\bibinfo {volume} {79}},\ \bibinfo {pages} {084003} (\bibinfo {year}
  {2009}{\natexlab{a}})},\ \Eprint {http://arxiv.org/abs/0901.1314}
  {arXiv:0901.1314 [hep-th]} \BibitemShut {NoStop}%
\bibitem [{\citenamefont {Deffayet}\ \emph
  {et~al.}(2009{\natexlab{b}})\citenamefont {Deffayet}, \citenamefont {Deser},\
  and\ \citenamefont {Esposito-Far{\`e}se}}]{Deffayet:2009mn}%
  \BibitemOpen
  \bibfield  {author} {\bibinfo {author} {\bibfnamefont {C.}~\bibnamefont
  {Deffayet}}, \bibinfo {author} {\bibfnamefont {S.}~\bibnamefont {Deser}},
  and \bibinfo {author} {\bibfnamefont {G.}~\bibnamefont
  {Esposito-Far{\`e}se}},\ }\href {\doibase 10.1103/PhysRevD.80.064015}
  {\bibfield  {journal} {\bibinfo  {journal} {\emph {Phys. Rev. D}}\ }\textbf
  {\bibinfo {volume} {80}},\ \bibinfo {pages} {064015} (\bibinfo {year}
  {2009}{\natexlab{b}})},\ \Eprint {http://arxiv.org/abs/0906.1967}
  {arXiv:0906.1967 [gr-qc]} \BibitemShut {NoStop}%
\bibitem [{\citenamefont {Deffayet}\ \emph {et~al.}(2011)\citenamefont
  {Deffayet}, \citenamefont {Gao}, \citenamefont {Steer},\ and\ \citenamefont
  {Zahariade}}]{Deffayet:2011gz}%
  \BibitemOpen
  \bibfield  {author} {\bibinfo {author} {\bibfnamefont {C.}~\bibnamefont
  {Deffayet}}, \bibinfo {author} {\bibfnamefont {X.}~\bibnamefont {Gao}},
  \bibinfo {author} {\bibfnamefont {D.~A.}\ \bibnamefont {Steer}},  and
  \bibinfo {author} {\bibfnamefont {G.}~\bibnamefont {Zahariade}},\ }\href
  {\doibase 10.1103/PhysRevD.84.064039} {\bibfield  {journal} {\bibinfo
  {journal} {\emph {Phys. Rev. D}}\ }\textbf {\bibinfo {volume} {84}},\
  \bibinfo {pages} {064039} (\bibinfo {year} {2011})},\ \Eprint
  {http://arxiv.org/abs/1103.3260} {arXiv:1103.3260 [hep-th]} \BibitemShut
  {NoStop}%
\bibitem [{\citenamefont {Kobayashi}\ \emph {et~al.}(2011)\citenamefont
  {Kobayashi}, \citenamefont {Yamaguchi},\ and\ \citenamefont
  {Yokoyama}}]{Kobayashi:2011nu}%
  \BibitemOpen
  \bibfield  {author} {\bibinfo {author} {\bibfnamefont {T.}~\bibnamefont
  {Kobayashi}}, \bibinfo {author} {\bibfnamefont {M.}~\bibnamefont
  {Yamaguchi}},  and \bibinfo {author} {\bibfnamefont {J.}~\bibnamefont
  {Yokoyama}},\ }\href {\doibase 10.1143/PTP.126.511} {\bibfield  {journal}
  {\bibinfo  {journal} {\emph {Prog. Theor. Phys.}}\ }\textbf {\bibinfo
  {volume} {126}},\ \bibinfo {pages} {511} (\bibinfo {year} {2011})},\ \Eprint
  {http://arxiv.org/abs/1105.5723} {arXiv:1105.5723 [hep-th]} \BibitemShut
  {NoStop}%
\bibitem [{\citenamefont {Gleyzes}\ \emph
  {et~al.}(2015{\natexlab{a}})\citenamefont {Gleyzes}, \citenamefont
  {Langlois}, \citenamefont {Piazza},\ and\ \citenamefont
  {Vernizzi}}]{Gleyzes:2014dya}%
  \BibitemOpen
  \bibfield  {author} {\bibinfo {author} {\bibfnamefont {J.}~\bibnamefont
  {Gleyzes}}, \bibinfo {author} {\bibfnamefont {D.}~\bibnamefont {Langlois}},
  \bibinfo {author} {\bibfnamefont {F.}~\bibnamefont {Piazza}},  and \bibinfo
  {author} {\bibfnamefont {F.}~\bibnamefont {Vernizzi}},\ }\href {\doibase
  10.1103/PhysRevLett.114.211101} {\bibfield  {journal} {\bibinfo  {journal}
  {\emph {Phys. Rev. Lett.}}\ }\textbf {\bibinfo {volume} {114}},\ \bibinfo
  {pages} {211101} (\bibinfo {year} {2015}{\natexlab{a}})},\ \Eprint
  {http://arxiv.org/abs/1404.6495} {arXiv:1404.6495 [hep-th]} \BibitemShut
  {NoStop}%
\bibitem [{\citenamefont {Gleyzes}\ \emph
  {et~al.}(2015{\natexlab{b}})\citenamefont {Gleyzes}, \citenamefont
  {Langlois}, \citenamefont {Piazza},\ and\ \citenamefont
  {Vernizzi}}]{Gleyzes:2014qga}%
  \BibitemOpen
  \bibfield  {author} {\bibinfo {author} {\bibfnamefont {J.}~\bibnamefont
  {Gleyzes}}, \bibinfo {author} {\bibfnamefont {D.}~\bibnamefont {Langlois}},
  \bibinfo {author} {\bibfnamefont {F.}~\bibnamefont {Piazza}},  and \bibinfo
  {author} {\bibfnamefont {F.}~\bibnamefont {Vernizzi}},\ }\href {\doibase
  10.1088/1475-7516/2015/02/018} {\bibfield  {journal} {\bibinfo  {journal}
  {\emph {JCAP}}\ }\textbf {\bibinfo {volume} {02}},\ \bibinfo {pages} {018}
  (\bibinfo {year} {2015}{\natexlab{b}})},\ \Eprint
  {http://arxiv.org/abs/1408.1952} {arXiv:1408.1952 [astro-ph.CO]} \BibitemShut
  {NoStop}%
\bibitem [{\citenamefont {Langlois}(2019)}]{Langlois:2018dxi}%
  \BibitemOpen
  \bibfield  {author} {\bibinfo {author} {\bibfnamefont {D.}~\bibnamefont
  {Langlois}},\ }\href {\doibase 10.1142/S0218271819420069} {\bibfield
  {journal} {\bibinfo  {journal} {\emph {Int. J. Mod. Phys. D}}\ }\textbf
  {\bibinfo {volume} {28}},\ \bibinfo {pages} {1942006} (\bibinfo {year}
  {2019})},\ \Eprint {http://arxiv.org/abs/1811.06271} {arXiv:1811.06271
  [gr-qc]} \BibitemShut {NoStop}%
\bibitem [{\citenamefont {Kobayashi}(2019)}]{Kobayashi:2019hrl}%
  \BibitemOpen
  \bibfield  {author} {\bibinfo {author} {\bibfnamefont {T.}~\bibnamefont
  {Kobayashi}},\ }\href {\doibase 10.1088/1361-6633/ab2429} {\bibfield
  {journal} {\bibinfo  {journal} {\emph {Rept.\ Prog.\ Phys.}}\ }\textbf
  {\bibinfo {volume} {82}},\ \bibinfo {pages} {086901} (\bibinfo {year}
  {2019})},\ \Eprint {http://arxiv.org/abs/1901.07183} {arXiv:1901.07183
  [gr-qc]} \BibitemShut {NoStop}%
\bibitem [{\citenamefont {Ben~Achour}\ and\ \citenamefont
  {Liu}(2019)}]{BenAchour:2018dap}%
  \BibitemOpen
  \bibfield  {author} {\bibinfo {author} {\bibfnamefont {J.}~\bibnamefont
  {Ben~Achour}} and \bibinfo {author} {\bibfnamefont {H.}~\bibnamefont {Liu}},\
  }\href {\doibase 10.1103/PhysRevD.99.064042} {\bibfield  {journal} {\bibinfo
  {journal} {\emph {Phys. Rev. D}}\ }\textbf {\bibinfo {volume} {99}},\
  \bibinfo {pages} {064042} (\bibinfo {year} {2019})},\ \Eprint
  {http://arxiv.org/abs/1811.05369} {arXiv:1811.05369 [gr-qc]} \BibitemShut
  {NoStop}%
\bibitem [{\citenamefont {Motohashi}\ and\ \citenamefont
  {Minamitsuji}(2019)}]{Motohashi:2019sen}%
  \BibitemOpen
  \bibfield  {author} {\bibinfo {author} {\bibfnamefont {H.}~\bibnamefont
  {Motohashi}} and \bibinfo {author} {\bibfnamefont {M.}~\bibnamefont
  {Minamitsuji}},\ }\href {\doibase 10.1103/PhysRevD.99.064040} {\bibfield
  {journal} {\bibinfo  {journal} {\emph {Phys. Rev. D}}\ }\textbf {\bibinfo
  {volume} {99}},\ \bibinfo {pages} {064040} (\bibinfo {year} {2019})},\
  \Eprint {http://arxiv.org/abs/1901.04658} {arXiv:1901.04658 [gr-qc]}
  \BibitemShut {NoStop}%
\bibitem [{\citenamefont {Minamitsuji}\ and\ \citenamefont
  {Edholm}(2019)}]{Minamitsuji:2019shy}%
  \BibitemOpen
  \bibfield  {author} {\bibinfo {author} {\bibfnamefont {M.}~\bibnamefont
  {Minamitsuji}} and \bibinfo {author} {\bibfnamefont {J.}~\bibnamefont
  {Edholm}},\ }\href {\doibase 10.1103/PhysRevD.100.044053} {\bibfield
  {journal} {\bibinfo  {journal} {\emph {Phys. Rev. D}}\ }\textbf {\bibinfo
  {volume} {100}},\ \bibinfo {pages} {044053} (\bibinfo {year} {2019})},\
  \Eprint {http://arxiv.org/abs/1907.02072} {arXiv:1907.02072 [gr-qc]}
  \BibitemShut {NoStop}%
\bibitem [{\citenamefont {Charmousis}\ \emph
  {et~al.}(2019{\natexlab{a}})\citenamefont {Charmousis}, \citenamefont
  {Crisostomi}, \citenamefont {Gregory},\ and\ \citenamefont
  {Stergioulas}}]{Charmousis:2019vnf}%
  \BibitemOpen
  \bibfield  {author} {\bibinfo {author} {\bibfnamefont {C.}~\bibnamefont
  {Charmousis}}, \bibinfo {author} {\bibfnamefont {M.}~\bibnamefont
  {Crisostomi}}, \bibinfo {author} {\bibfnamefont {R.}~\bibnamefont {Gregory}},
   and \bibinfo {author} {\bibfnamefont {N.}~\bibnamefont {Stergioulas}},\
  }\href {\doibase 10.1103/PhysRevD.100.084020} {\bibfield  {journal} {\bibinfo
   {journal} {\emph {Phys. Rev. D}}\ }\textbf {\bibinfo {volume} {100}},\
  \bibinfo {pages} {084020} (\bibinfo {year} {2019}{\natexlab{a}})},\ \Eprint
  {http://arxiv.org/abs/1903.05519} {arXiv:1903.05519 [hep-th]} \BibitemShut
  {NoStop}%
\bibitem [{\citenamefont {Babichev}\ and\ \citenamefont
  {Charmousis}(2014)}]{Babichev:2013cya}%
  \BibitemOpen
  \bibfield  {author} {\bibinfo {author} {\bibfnamefont {E.}~\bibnamefont
  {Babichev}} and \bibinfo {author} {\bibfnamefont {C.}~\bibnamefont
  {Charmousis}},\ }\href {\doibase 10.1007/JHEP08(2014)106} {\bibfield
  {journal} {\bibinfo  {journal} {\emph {JHEP}}\ }\textbf {\bibinfo {volume}
  {08}},\ \bibinfo {pages} {106} (\bibinfo {year} {2014})},\ \Eprint
  {http://arxiv.org/abs/1312.3204} {arXiv:1312.3204 [gr-qc]} \BibitemShut
  {NoStop}%
\bibitem [{\citenamefont {Kobayashi}\ and\ \citenamefont
  {Tanahashi}(2014)}]{Kobayashi:2014eva}%
  \BibitemOpen
  \bibfield  {author} {\bibinfo {author} {\bibfnamefont {T.}~\bibnamefont
  {Kobayashi}} and \bibinfo {author} {\bibfnamefont {N.}~\bibnamefont
  {Tanahashi}},\ }\href {\doibase 10.1093/ptep/ptu096} {\bibfield  {journal}
  {\bibinfo  {journal} {\emph {PTEP}}\ }\textbf {\bibinfo {volume} {2014}},\
  \bibinfo {pages} {073E02} (\bibinfo {year} {2014})},\ \Eprint
  {http://arxiv.org/abs/1403.4364} {arXiv:1403.4364 [gr-qc]} \BibitemShut
  {NoStop}%
\bibitem [{\citenamefont {Babichev}\ and\ \citenamefont
  {Esposito-Far{\`e}se}(2017)}]{Babichev:2016kdt}%
  \BibitemOpen
  \bibfield  {author} {\bibinfo {author} {\bibfnamefont {E.}~\bibnamefont
  {Babichev}} and \bibinfo {author} {\bibfnamefont {G.}~\bibnamefont
  {Esposito-Far{\`e}se}},\ }\href {\doibase 10.1103/PhysRevD.95.024020}
  {\bibfield  {journal} {\bibinfo  {journal} {\emph {Phys. Rev. D}}\ }\textbf
  {\bibinfo {volume} {95}},\ \bibinfo {pages} {024020} (\bibinfo {year}
  {2017})},\ \Eprint {http://arxiv.org/abs/1609.09798} {arXiv:1609.09798
  [gr-qc]} \BibitemShut {NoStop}%
\bibitem [{\citenamefont {Babichev}\ \emph {et~al.}(2017)\citenamefont
  {Babichev}, \citenamefont {Charmousis},\ and\ \citenamefont
  {Leh\'ebel}}]{Babichev:2017guv}%
  \BibitemOpen
  \bibfield  {author} {\bibinfo {author} {\bibfnamefont {E.}~\bibnamefont
  {Babichev}}, \bibinfo {author} {\bibfnamefont {C.}~\bibnamefont
  {Charmousis}},  and \bibinfo {author} {\bibfnamefont {A.}~\bibnamefont
  {Leh\'ebel}},\ }\href {\doibase 10.1088/1475-7516/2017/04/027} {\bibfield
  {journal} {\bibinfo  {journal} {\emph {JCAP}}\ }\textbf {\bibinfo {volume}
  {04}},\ \bibinfo {pages} {027} (\bibinfo {year} {2017})},\ \Eprint
  {http://arxiv.org/abs/1702.01938} {arXiv:1702.01938 [gr-qc]} \BibitemShut
  {NoStop}%
\bibitem [{\citenamefont {Babichev}\ \emph
  {et~al.}(2018{\natexlab{a}})\citenamefont {Babichev}, \citenamefont
  {Charmousis}, \citenamefont {Esposito-Far\`ese},\ and\ \citenamefont
  {Leh\'ebel}}]{Babichev:2017lmw}%
  \BibitemOpen
  \bibfield  {author} {\bibinfo {author} {\bibfnamefont {E.}~\bibnamefont
  {Babichev}}, \bibinfo {author} {\bibfnamefont {C.}~\bibnamefont
  {Charmousis}}, \bibinfo {author} {\bibfnamefont {G.}~\bibnamefont
  {Esposito-Far\`ese}},  and \bibinfo {author} {\bibfnamefont {A.}~\bibnamefont
  {Leh\'ebel}},\ }\href {\doibase 10.1103/PhysRevLett.120.241101} {\bibfield
  {journal} {\bibinfo  {journal} {\emph {Phys. Rev. Lett.}}\ }\textbf {\bibinfo
  {volume} {120}},\ \bibinfo {pages} {241101} (\bibinfo {year}
  {2018}{\natexlab{a}})},\ \Eprint {http://arxiv.org/abs/1712.04398}
  {arXiv:1712.04398 [gr-qc]} \BibitemShut {NoStop}%
\bibitem [{\citenamefont {Ogawa}\ \emph {et~al.}(2016)\citenamefont {Ogawa},
  \citenamefont {Kobayashi},\ and\ \citenamefont {Suyama}}]{Ogawa:2015pea}%
  \BibitemOpen
  \bibfield  {author} {\bibinfo {author} {\bibfnamefont {H.}~\bibnamefont
  {Ogawa}}, \bibinfo {author} {\bibfnamefont {T.}~\bibnamefont {Kobayashi}},
  and \bibinfo {author} {\bibfnamefont {T.}~\bibnamefont {Suyama}},\ }\href
  {\doibase 10.1103/PhysRevD.93.064078} {\bibfield  {journal} {\bibinfo
  {journal} {\emph {Phys. Rev. D}}\ }\textbf {\bibinfo {volume} {93}},\
  \bibinfo {pages} {064078} (\bibinfo {year} {2016})},\ \Eprint
  {http://arxiv.org/abs/1510.07400} {arXiv:1510.07400 [gr-qc]} \BibitemShut
  {NoStop}%
\bibitem [{\citenamefont {Takahashi}\ \emph {et~al.}(2016)\citenamefont
  {Takahashi}, \citenamefont {Suyama},\ and\ \citenamefont
  {Kobayashi}}]{Takahashi:2015pad}%
  \BibitemOpen
  \bibfield  {author} {\bibinfo {author} {\bibfnamefont {K.}~\bibnamefont
  {Takahashi}}, \bibinfo {author} {\bibfnamefont {T.}~\bibnamefont {Suyama}},
  and \bibinfo {author} {\bibfnamefont {T.}~\bibnamefont {Kobayashi}},\ }\href
  {\doibase 10.1103/PhysRevD.93.064068} {\bibfield  {journal} {\bibinfo
  {journal} {\emph {Phys. Rev. D}}\ }\textbf {\bibinfo {volume} {93}},\
  \bibinfo {pages} {064068} (\bibinfo {year} {2016})},\ \Eprint
  {http://arxiv.org/abs/1511.06083} {arXiv:1511.06083 [gr-qc]} \BibitemShut
  {NoStop}%
\bibitem [{\citenamefont {Takahashi}\ and\ \citenamefont
  {Suyama}(2017)}]{Takahashi:2016dnv}%
  \BibitemOpen
  \bibfield  {author} {\bibinfo {author} {\bibfnamefont {K.}~\bibnamefont
  {Takahashi}} and \bibinfo {author} {\bibfnamefont {T.}~\bibnamefont
  {Suyama}},\ }\href {\doibase 10.1103/PhysRevD.95.024034} {\bibfield
  {journal} {\bibinfo  {journal} {\emph {Phys. Rev. D}}\ }\textbf {\bibinfo
  {volume} {95}},\ \bibinfo {pages} {024034} (\bibinfo {year} {2017})},\
  \Eprint {http://arxiv.org/abs/1610.00432} {arXiv:1610.00432 [gr-qc]}
  \BibitemShut {NoStop}%
\bibitem [{\citenamefont {Tretyakova}\ and\ \citenamefont
  {Takahashi}(2017)}]{Tretyakova:2017lyg}%
  \BibitemOpen
  \bibfield  {author} {\bibinfo {author} {\bibfnamefont {D.~A.}\ \bibnamefont
  {Tretyakova}} and \bibinfo {author} {\bibfnamefont {K.}~\bibnamefont
  {Takahashi}},\ }\href {\doibase 10.1088/1361-6382/aa8057} {\bibfield
  {journal} {\bibinfo  {journal} {\emph {Class. Quant. Grav.}}\ }\textbf
  {\bibinfo {volume} {34}},\ \bibinfo {pages} {175007} (\bibinfo {year}
  {2017})},\ \Eprint {http://arxiv.org/abs/1702.03502} {arXiv:1702.03502
  [gr-qc]} \BibitemShut {NoStop}%
\bibitem [{\citenamefont {Babichev}\ \emph
  {et~al.}(2018{\natexlab{b}})\citenamefont {Babichev}, \citenamefont
  {Charmousis}, \citenamefont {Esposito-Far\`ese},\ and\ \citenamefont
  {Leh\'ebel}}]{Babichev:2018uiw}%
  \BibitemOpen
  \bibfield  {author} {\bibinfo {author} {\bibfnamefont {E.}~\bibnamefont
  {Babichev}}, \bibinfo {author} {\bibfnamefont {C.}~\bibnamefont
  {Charmousis}}, \bibinfo {author} {\bibfnamefont {G.}~\bibnamefont
  {Esposito-Far\`ese}},  and \bibinfo {author} {\bibfnamefont {A.}~\bibnamefont
  {Leh\'ebel}},\ }\href {\doibase 10.1103/PhysRevD.98.104050} {\bibfield
  {journal} {\bibinfo  {journal} {\emph {Phys. Rev. D}}\ }\textbf {\bibinfo
  {volume} {98}},\ \bibinfo {pages} {104050} (\bibinfo {year}
  {2018}{\natexlab{b}})},\ \Eprint {http://arxiv.org/abs/1803.11444}
  {arXiv:1803.11444 [gr-qc]} \BibitemShut {NoStop}%
\bibitem [{\citenamefont {Takahashi}\ \emph {et~al.}(2019)\citenamefont
  {Takahashi}, \citenamefont {Motohashi},\ and\ \citenamefont
  {Minamitsuji}}]{Takahashi:2019oxz}%
  \BibitemOpen
  \bibfield  {author} {\bibinfo {author} {\bibfnamefont {K.}~\bibnamefont
  {Takahashi}}, \bibinfo {author} {\bibfnamefont {H.}~\bibnamefont
  {Motohashi}},  and \bibinfo {author} {\bibfnamefont {M.}~\bibnamefont
  {Minamitsuji}},\ }\href {\doibase 10.1103/PhysRevD.100.024041} {\bibfield
  {journal} {\bibinfo  {journal} {\emph {Phys. Rev. D}}\ }\textbf {\bibinfo
  {volume} {100}},\ \bibinfo {pages} {024041} (\bibinfo {year} {2019})},\
  \Eprint {http://arxiv.org/abs/1904.03554} {arXiv:1904.03554 [gr-qc]}
  \BibitemShut {NoStop}%
\bibitem [{\citenamefont {de~Rham}\ and\ \citenamefont
  {Zhang}(2019)}]{deRham:2019slh}%
  \BibitemOpen
  \bibfield  {author} {\bibinfo {author} {\bibfnamefont {C.}~\bibnamefont
  {de~Rham}} and \bibinfo {author} {\bibfnamefont {J.}~\bibnamefont {Zhang}},\
  }\href {\doibase 10.1103/PhysRevD.100.124023} {\bibfield  {journal} {\bibinfo
   {journal} {\emph {Phys. Rev. D}}\ }\textbf {\bibinfo {volume} {100}},\
  \bibinfo {pages} {124023} (\bibinfo {year} {2019})},\ \Eprint
  {http://arxiv.org/abs/1907.00699} {arXiv:1907.00699 [hep-th]} \BibitemShut
  {NoStop}%
\bibitem [{\citenamefont {Charmousis}\ \emph
  {et~al.}(2019{\natexlab{b}})\citenamefont {Charmousis}, \citenamefont
  {Crisostomi}, \citenamefont {Langlois},\ and\ \citenamefont
  {Noui}}]{Charmousis:2019fre}%
  \BibitemOpen
  \bibfield  {author} {\bibinfo {author} {\bibfnamefont {C.}~\bibnamefont
  {Charmousis}}, \bibinfo {author} {\bibfnamefont {M.}~\bibnamefont
  {Crisostomi}}, \bibinfo {author} {\bibfnamefont {D.}~\bibnamefont
  {Langlois}},  and \bibinfo {author} {\bibfnamefont {K.}~\bibnamefont
  {Noui}},\ }\href {\doibase 10.1088/1361-6382/ab4fb1} {\bibfield  {journal}
  {\bibinfo  {journal} {\emph {Class. Quant. Grav.}}\ }\textbf {\bibinfo
  {volume} {36}},\ \bibinfo {pages} {235008} (\bibinfo {year}
  {2019}{\natexlab{b}})},\ \Eprint {http://arxiv.org/abs/1907.02924}
  {arXiv:1907.02924 [gr-qc]} \BibitemShut {NoStop}%
\bibitem [{\citenamefont {Motohashi}\ and\ \citenamefont
  {Mukohyama}(2020)}]{Motohashi:2019ymr}%
  \BibitemOpen
  \bibfield  {author} {\bibinfo {author} {\bibfnamefont {H.}~\bibnamefont
  {Motohashi}} and \bibinfo {author} {\bibfnamefont {S.}~\bibnamefont
  {Mukohyama}},\ }\href {\doibase 10.1088/1475-7516/2020/01/030} {\bibfield
  {journal} {\bibinfo  {journal} {\emph {JCAP}}\ }\textbf {\bibinfo {volume}
  {01}},\ \bibinfo {pages} {030} (\bibinfo {year} {2020})},\ \Eprint
  {http://arxiv.org/abs/1912.00378} {arXiv:1912.00378 [gr-qc]} \BibitemShut
  {NoStop}%
\bibitem [{\citenamefont {Minamitsuji}\ and\ \citenamefont
  {Motohashi}(2018)}]{Minamitsuji:2018vuw}%
  \BibitemOpen
  \bibfield  {author} {\bibinfo {author} {\bibfnamefont {M.}~\bibnamefont
  {Minamitsuji}} and \bibinfo {author} {\bibfnamefont {H.}~\bibnamefont
  {Motohashi}},\ }\href {\doibase 10.1103/PhysRevD.98.084027} {\bibfield
  {journal} {\bibinfo  {journal} {\emph {Phys. Rev. D}}\ }\textbf {\bibinfo
  {volume} {98}},\ \bibinfo {pages} {084027} (\bibinfo {year} {2018})},\
  \Eprint {http://arxiv.org/abs/1809.06611} {arXiv:1809.06611 [gr-qc]}
  \BibitemShut {NoStop}%
\bibitem [{\citenamefont {Babichev}\ \emph {et~al.}(2015)\citenamefont
  {Babichev}, \citenamefont {Charmousis},\ and\ \citenamefont
  {Hassaine}}]{Babichev:2015rva}%
  \BibitemOpen
  \bibfield  {author} {\bibinfo {author} {\bibfnamefont {E.}~\bibnamefont
  {Babichev}}, \bibinfo {author} {\bibfnamefont {C.}~\bibnamefont
  {Charmousis}},  and \bibinfo {author} {\bibfnamefont {M.}~\bibnamefont
  {Hassaine}},\ }\href {\doibase 10.1088/1475-7516/2015/05/031} {\bibfield
  {journal} {\bibinfo  {journal} {\emph {JCAP}}\ }\textbf {\bibinfo {volume}
  {05}},\ \bibinfo {pages} {031} (\bibinfo {year} {2015})},\ \Eprint
  {http://arxiv.org/abs/1503.02545} {arXiv:1503.02545 [gr-qc]} \BibitemShut
  {NoStop}%
\bibitem [{\citenamefont {Ben~Achour}\ \emph
  {et~al.}(2016{\natexlab{b}})\citenamefont {Ben~Achour}, \citenamefont
  {Langlois},\ and\ \citenamefont {Noui}}]{Achour:2016rkg}%
  \BibitemOpen
  \bibfield  {author} {\bibinfo {author} {\bibfnamefont {J.}~\bibnamefont
  {Ben~Achour}}, \bibinfo {author} {\bibfnamefont {D.}~\bibnamefont
  {Langlois}},  and \bibinfo {author} {\bibfnamefont {K.}~\bibnamefont
  {Noui}},\ }\href {\doibase 10.1103/PhysRevD.93.124005} {\bibfield  {journal}
  {\bibinfo  {journal} {\emph {Phys. Rev. D}}\ }\textbf {\bibinfo {volume}
  {93}},\ \bibinfo {pages} {124005} (\bibinfo {year} {2016}{\natexlab{b}})},\
  \Eprint {http://arxiv.org/abs/1602.08398} {arXiv:1602.08398 [gr-qc]}
  \BibitemShut {NoStop}%
\bibitem [{\citenamefont {Motohashi}\ \emph
  {et~al.}(2016{\natexlab{b}})\citenamefont {Motohashi}, \citenamefont
  {Suyama},\ and\ \citenamefont {Takahashi}}]{Motohashi:2016prk}%
  \BibitemOpen
  \bibfield  {author} {\bibinfo {author} {\bibfnamefont {H.}~\bibnamefont
  {Motohashi}}, \bibinfo {author} {\bibfnamefont {T.}~\bibnamefont {Suyama}},
  and \bibinfo {author} {\bibfnamefont {K.}~\bibnamefont {Takahashi}},\ }\href
  {\doibase 10.1103/PhysRevD.94.124021} {\bibfield  {journal} {\bibinfo
  {journal} {\emph {Phys. Rev. D}}\ }\textbf {\bibinfo {volume} {94}},\
  \bibinfo {pages} {124021} (\bibinfo {year} {2016}{\natexlab{b}})},\ \Eprint
  {http://arxiv.org/abs/1608.00071} {arXiv:1608.00071 [gr-qc]} \BibitemShut
  {NoStop}%
\bibitem [{\citenamefont {Bravo-Gaete}\ and\ \citenamefont
  {Hassaine}(2014)}]{Bravo-Gaete:2014haa}%
  \BibitemOpen
  \bibfield  {author} {\bibinfo {author} {\bibfnamefont {M.}~\bibnamefont
  {Bravo-Gaete}} and \bibinfo {author} {\bibfnamefont {M.}~\bibnamefont
  {Hassaine}},\ }\href {\doibase 10.1103/PhysRevD.90.024008} {\bibfield
  {journal} {\bibinfo  {journal} {\emph {Phys. Rev. D}}\ }\textbf {\bibinfo
  {volume} {90}},\ \bibinfo {pages} {024008} (\bibinfo {year} {2014})},\
  \Eprint {http://arxiv.org/abs/1405.4935} {arXiv:1405.4935 [hep-th]}
  \BibitemShut {NoStop}%
\bibitem [{\citenamefont {Kobayashi}\ \emph {et~al.}(2014)\citenamefont
  {Kobayashi}, \citenamefont {Motohashi},\ and\ \citenamefont
  {Suyama}}]{Kobayashi:2014wsa}%
  \BibitemOpen
  \bibfield  {author} {\bibinfo {author} {\bibfnamefont {T.}~\bibnamefont
  {Kobayashi}}, \bibinfo {author} {\bibfnamefont {H.}~\bibnamefont
  {Motohashi}},  and \bibinfo {author} {\bibfnamefont {T.}~\bibnamefont
  {Suyama}},\ }\href {\doibase 10.1103/PhysRevD.89.084042} {\bibfield
  {journal} {\bibinfo  {journal} {\emph {Phys. Rev. D}}\ }\textbf {\bibinfo
  {volume} {89}},\ \bibinfo {pages} {084042} (\bibinfo {year} {2014})},\
  \Eprint {http://arxiv.org/abs/1402.6740} {arXiv:1402.6740 [gr-qc]}
  \BibitemShut {NoStop}%
\bibitem [{\citenamefont {Mironov}\ \emph {et~al.}(2019)\citenamefont
  {Mironov}, \citenamefont {Rubakov},\ and\ \citenamefont
  {Volkova}}]{Mironov:2018uou}%
  \BibitemOpen
  \bibfield  {author} {\bibinfo {author} {\bibfnamefont {S.}~\bibnamefont
  {Mironov}}, \bibinfo {author} {\bibfnamefont {V.}~\bibnamefont {Rubakov}},
  and \bibinfo {author} {\bibfnamefont {V.}~\bibnamefont {Volkova}},\ }\href
  {\doibase 10.1088/1361-6382/ab2574} {\bibfield  {journal} {\bibinfo
  {journal} {\emph {Class. Quant. Grav.}}\ }\textbf {\bibinfo {volume} {36}},\
  \bibinfo {pages} {135008} (\bibinfo {year} {2019})},\ \Eprint
  {http://arxiv.org/abs/1812.07022} {arXiv:1812.07022 [hep-th]} \BibitemShut
  {NoStop}%
\bibitem [{\citenamefont {Carter}(1970)}]{Carter:1970ea}%
  \BibitemOpen
  \bibfield  {author} {\bibinfo {author} {\bibfnamefont {B.}~\bibnamefont
  {Carter}},\ }\href {\doibase 10.1007/BF01647092} {\bibfield  {journal}
  {\bibinfo  {journal} {\emph {Commun. Math. Phys.}}\ }\textbf {\bibinfo
  {volume} {17}},\ \bibinfo {pages} {233} (\bibinfo {year} {1970})}\BibitemShut
  {NoStop}%
\bibitem [{\citenamefont {Carter}(1973)}]{Carter:1973rla}%
  \BibitemOpen
  \bibfield  {author} {\bibinfo {author} {\bibfnamefont {B.}~\bibnamefont
  {Carter}},\ }in\ \href@noop {} {\emph {\bibinfo {booktitle} {{Proceedings,
  {\'E}cole d'{\'E}t{\'e} de Physique Th{\'e}orique: Les Astres Occlus: Les
  Houches, France, August, 1972}}}}\ (\bibinfo {year} {1973})\ pp.\ \bibinfo
  {pages} {57--214}\BibitemShut {NoStop}%
\bibitem [{\citenamefont {Carter}(1968)}]{Carter:1968rr}%
  \BibitemOpen
  \bibfield  {author} {\bibinfo {author} {\bibfnamefont {B.}~\bibnamefont
  {Carter}},\ }\href {\doibase 10.1103/PhysRev.174.1559} {\bibfield  {journal}
  {\bibinfo  {journal} {\emph {Phys. Rev.}}\ }\textbf {\bibinfo {volume}
  {174}},\ \bibinfo {pages} {1559} (\bibinfo {year} {1968})}\BibitemShut
  {NoStop}%
\bibitem [{\citenamefont {Ba{\~n}ados}\ \emph {et~al.}(1992)\citenamefont
  {Ba{\~n}ados}, \citenamefont {Teitelboim},\ and\ \citenamefont
  {Zanelli}}]{Banados:1992wn}%
  \BibitemOpen
  \bibfield  {author} {\bibinfo {author} {\bibfnamefont {M.}~\bibnamefont
  {Ba{\~n}ados}}, \bibinfo {author} {\bibfnamefont {C.}~\bibnamefont
  {Teitelboim}},  and \bibinfo {author} {\bibfnamefont {J.}~\bibnamefont
  {Zanelli}},\ }\href {\doibase 10.1103/PhysRevLett.69.1849} {\bibfield
  {journal} {\bibinfo  {journal} {\emph {Phys.\ Rev.\ Lett.}}\ }\textbf
  {\bibinfo {volume} {69}},\ \bibinfo {pages} {1849} (\bibinfo {year}
  {1992})},\ \Eprint {http://arxiv.org/abs/hep-th/9204099}
  {arXiv:hep-th/9204099} \BibitemShut {NoStop}%
\bibitem [{\citenamefont {Oppenheimer}\ and\ \citenamefont
  {Snyder}(1939)}]{Oppenheimer:1939ue}%
  \BibitemOpen
  \bibfield  {author} {\bibinfo {author} {\bibfnamefont {J.~R.}\ \bibnamefont
  {Oppenheimer}} and \bibinfo {author} {\bibfnamefont {H.}~\bibnamefont
  {Snyder}},\ }\href {\doibase 10.1103/PhysRev.56.455} {\bibfield  {journal}
  {\bibinfo  {journal} {\emph {Phys. Rev.}}\ }\textbf {\bibinfo {volume}
  {56}},\ \bibinfo {pages} {455} (\bibinfo {year} {1939})}\BibitemShut
  {NoStop}%
\bibitem [{\citenamefont {Nakao}(1991)}]{Nakao:1991sh}%
  \BibitemOpen
  \bibfield  {author} {\bibinfo {author} {\bibfnamefont {K.}~\bibnamefont
  {Nakao}},\ }\href {\doibase 10.1007/BF00756947} {\bibfield  {journal}
  {\bibinfo  {journal} {\emph {Gen. Rel. Grav.}}\ }\textbf {\bibinfo {volume}
  {24}},\ \bibinfo {pages} {1069–1081} (\bibinfo {year} {1991})}\BibitemShut
  {NoStop}%
\bibitem [{\citenamefont {Pleba{\'n}ski}\ and\ \citenamefont
  {Demia{\'n}ski}(1976)}]{Plebanski:1976gy}%
  \BibitemOpen
  \bibfield  {author} {\bibinfo {author} {\bibfnamefont {J.~F.}\ \bibnamefont
  {Pleba{\'n}ski}} and \bibinfo {author} {\bibfnamefont {M.}~\bibnamefont
  {Demia{\'n}ski}},\ }\href {\doibase 10.1016/0003-4916(76)90240-2} {\bibfield
  {journal} {\bibinfo  {journal} {\emph {Annals Phys.}}\ }\textbf {\bibinfo
  {volume} {98}},\ \bibinfo {pages} {98} (\bibinfo {year} {1976})}\BibitemShut
  {NoStop}%
\bibitem [{\citenamefont {Lin}\ and\ \citenamefont
  {Mukohyama}(2017)}]{Lin:2017oow}%
  \BibitemOpen
  \bibfield  {author} {\bibinfo {author} {\bibfnamefont {C.}~\bibnamefont
  {Lin}} and \bibinfo {author} {\bibfnamefont {S.}~\bibnamefont {Mukohyama}},\
  }\href {\doibase 10.1088/1475-7516/2017/10/033} {\bibfield  {journal}
  {\bibinfo  {journal} {\emph {JCAP}}\ }\textbf {\bibinfo {volume} {10}},\
  \bibinfo {pages} {033} (\bibinfo {year} {2017})},\ \Eprint
  {http://arxiv.org/abs/1708.03757} {arXiv:1708.03757 [gr-qc]} \BibitemShut
  {NoStop}%
\bibitem [{\citenamefont {Chagoya}\ and\ \citenamefont
  {Tasinato}(2019)}]{Chagoya:2018yna}%
  \BibitemOpen
  \bibfield  {author} {\bibinfo {author} {\bibfnamefont {J.}~\bibnamefont
  {Chagoya}} and \bibinfo {author} {\bibfnamefont {G.}~\bibnamefont
  {Tasinato}},\ }\href {\doibase 10.1088/1361-6382/ab0a4b} {\bibfield
  {journal} {\bibinfo  {journal} {\emph {Class. Quant. Grav.}}\ }\textbf
  {\bibinfo {volume} {36}},\ \bibinfo {pages} {075014} (\bibinfo {year}
  {2019})},\ \Eprint {http://arxiv.org/abs/1805.12010} {arXiv:1805.12010
  [hep-th]} \BibitemShut {NoStop}%
\bibitem [{\citenamefont {Aoki}\ \emph {et~al.}(2018)\citenamefont {Aoki},
  \citenamefont {Lin},\ and\ \citenamefont {Mukohyama}}]{Aoki:2018zcv}%
  \BibitemOpen
  \bibfield  {author} {\bibinfo {author} {\bibfnamefont {K.}~\bibnamefont
  {Aoki}}, \bibinfo {author} {\bibfnamefont {C.}~\bibnamefont {Lin}},  and
  \bibinfo {author} {\bibfnamefont {S.}~\bibnamefont {Mukohyama}},\ }\href
  {\doibase 10.1103/PhysRevD.98.044022} {\bibfield  {journal} {\bibinfo
  {journal} {\emph {Phys. Rev. D}}\ }\textbf {\bibinfo {volume} {98}},\
  \bibinfo {pages} {044022} (\bibinfo {year} {2018})},\ \Eprint
  {http://arxiv.org/abs/1804.03902} {arXiv:1804.03902 [gr-qc]} \BibitemShut
  {NoStop}%
\bibitem [{\citenamefont {Afshordi}\ \emph {et~al.}(2007)\citenamefont
  {Afshordi}, \citenamefont {Chung},\ and\ \citenamefont
  {Geshnizjani}}]{Afshordi:2006ad}%
  \BibitemOpen
  \bibfield  {author} {\bibinfo {author} {\bibfnamefont {N.}~\bibnamefont
  {Afshordi}}, \bibinfo {author} {\bibfnamefont {D.~J.~H.}\ \bibnamefont
  {Chung}},  and \bibinfo {author} {\bibfnamefont {G.}~\bibnamefont
  {Geshnizjani}},\ }\href {\doibase 10.1103/PhysRevD.75.083513} {\bibfield
  {journal} {\bibinfo  {journal} {\emph {Phys. Rev. D}}\ }\textbf {\bibinfo
  {volume} {75}},\ \bibinfo {pages} {083513} (\bibinfo {year} {2007})},\
  \Eprint {http://arxiv.org/abs/hep-th/0609150} {arXiv:hep-th/0609150 [hep-th]}
  \BibitemShut {NoStop}%
\bibitem [{\citenamefont {Iyonaga}\ \emph {et~al.}(2018)\citenamefont
  {Iyonaga}, \citenamefont {Takahashi},\ and\ \citenamefont
  {Kobayashi}}]{Iyonaga:2018vnu}%
  \BibitemOpen
  \bibfield  {author} {\bibinfo {author} {\bibfnamefont {A.}~\bibnamefont
  {Iyonaga}}, \bibinfo {author} {\bibfnamefont {K.}~\bibnamefont {Takahashi}},
  and \bibinfo {author} {\bibfnamefont {T.}~\bibnamefont {Kobayashi}},\ }\href
  {\doibase 10.1088/1475-7516/2018/12/002} {\bibfield  {journal} {\bibinfo
  {journal} {\emph {JCAP}}\ }\textbf {\bibinfo {volume} {12}},\ \bibinfo
  {pages} {002} (\bibinfo {year} {2018})},\ \Eprint
  {http://arxiv.org/abs/1809.10935} {arXiv:1809.10935 [gr-qc]} \BibitemShut
  {NoStop}%
\bibitem [{\citenamefont {Gao}\ and\ \citenamefont {Yao}(2020)}]{Gao:2019twq}%
  \BibitemOpen
  \bibfield  {author} {\bibinfo {author} {\bibfnamefont {X.}~\bibnamefont
  {Gao}} and \bibinfo {author} {\bibfnamefont {Z.-B.}\ \bibnamefont {Yao}},\
  }\href {\doibase 10.1103/PhysRevD.101.064018} {\bibfield  {journal} {\bibinfo
   {journal} {\emph {Phys. Rev. D}}\ }\textbf {\bibinfo {volume} {101}},\
  \bibinfo {pages} {064018} (\bibinfo {year} {2020})},\ \Eprint
  {http://arxiv.org/abs/1910.13995} {arXiv:1910.13995 [gr-qc]} \BibitemShut
  {NoStop}%
\bibitem [{\citenamefont {Ben~Achour}\ \emph
  {et~al.}(2020{\natexlab{a}})\citenamefont {Ben~Achour}, \citenamefont {Liu},\
  and\ \citenamefont {Mukohyama}}]{BenAchour:2019fdf}%
  \BibitemOpen
  \bibfield  {author} {\bibinfo {author} {\bibfnamefont {J.}~\bibnamefont
  {Ben~Achour}}, \bibinfo {author} {\bibfnamefont {H.}~\bibnamefont {Liu}},
  and \bibinfo {author} {\bibfnamefont {S.}~\bibnamefont {Mukohyama}},\ }\href
  {\doibase 10.1088/1475-7516/2020/02/023} {\bibfield  {journal} {\bibinfo
  {journal} {\emph {JCAP}}\ }\textbf {\bibinfo {volume} {02}},\ \bibinfo
  {pages} {023} (\bibinfo {year} {2020}{\natexlab{a}})},\ \Eprint
  {http://arxiv.org/abs/1910.11017} {arXiv:1910.11017 [gr-qc]} \BibitemShut
  {NoStop}%
\bibitem [{\citenamefont {Babichev}\ \emph {et~al.}(2020)\citenamefont
  {Babichev}, \citenamefont {Charmousis}, \citenamefont {Cisterna},\ and\
  \citenamefont {Hassaine}}]{Babichev:2020qpr}%
  \BibitemOpen
  \bibfield  {author} {\bibinfo {author} {\bibfnamefont {E.}~\bibnamefont
  {Babichev}}, \bibinfo {author} {\bibfnamefont {C.}~\bibnamefont
  {Charmousis}}, \bibinfo {author} {\bibfnamefont {A.}~\bibnamefont
  {Cisterna}},  and \bibinfo {author} {\bibfnamefont {M.}~\bibnamefont
  {Hassaine}},\ }\Eprint {http://arxiv.org/abs/2004.00597} {arXiv:2004.00597
  [hep-th]} \BibitemShut {NoStop}%
\bibitem [{\citenamefont {Ben~Achour}\ \emph
  {et~al.}(2020{\natexlab{b}})\citenamefont {Ben~Achour}, \citenamefont {Liu},
  \citenamefont {Motohashi}, \citenamefont {Mukohyama},\ and\ \citenamefont
  {Noui}}]{BenAchour:2020fgy}%
  \BibitemOpen
  \bibfield  {author} {\bibinfo {author} {\bibfnamefont {J.}~\bibnamefont
  {Ben~Achour}}, \bibinfo {author} {\bibfnamefont {H.}~\bibnamefont {Liu}},
  \bibinfo {author} {\bibfnamefont {H.}~\bibnamefont {Motohashi}}, \bibinfo
  {author} {\bibfnamefont {S.}~\bibnamefont {Mukohyama}},  and \bibinfo
  {author} {\bibfnamefont {K.}~\bibnamefont {Noui}},\ }\Eprint
  {http://arxiv.org/abs/2006.07245} {arXiv:2006.07245 [gr-qc]} \BibitemShut
  {NoStop}%
\bibitem [{\citenamefont {Takahashi}\ and\ \citenamefont
  {Kobayashi}(2019)}]{Takahashi:2018yzc}%
  \BibitemOpen
  \bibfield  {author} {\bibinfo {author} {\bibfnamefont {K.}~\bibnamefont
  {Takahashi}} and \bibinfo {author} {\bibfnamefont {T.}~\bibnamefont
  {Kobayashi}},\ }\href {\doibase 10.1088/1361-6382/ab1355} {\bibfield
  {journal} {\bibinfo  {journal} {\emph {Class.\ Quant.\ Grav.}}\ }\textbf
  {\bibinfo {volume} {36}},\ \bibinfo {pages} {095003} (\bibinfo {year}
  {2019})},\ \Eprint {http://arxiv.org/abs/1812.08847} {arXiv:1812.08847
  [gr-qc]} \BibitemShut {NoStop}%
\end{thebibliography}%

\end{document}